\documentclass[11pt,letter]{llncs}

\usepackage[utf8]{inputenc}
\usepackage{graphicx}
\usepackage{hyperref}
\usepackage{multirow}
\usepackage{forest}
\usepackage{booktabs}
\usepackage{xspace}
\usepackage{amsmath, amsfonts,amssymb}
\usepackage{todonotes}
\usepackage{subcaption}
\usepackage{tikz}
\usepackage{relsize}
\usepackage[margin=1in]{geometry}
\usepackage{dirtytalk}
\usepackage{dsfont}
\usepackage{subcaption}
\usepackage{comment}
\usepackage[ruled,vlined,linesnumbered,noresetcount]{algorithm2e}
\usepackage{paracol}
\usepackage{wrapfig}
\usepackage{float}
\usepackage{tabularx}
\newcolumntype{C}{>{\centering\arraybackslash}X}

\usepackage{placeins}

\let\llncssubparagraph\subparagraph
\let\subparagraph\paragraph
\usepackage{titlesec}
\let\subparagraph\llncssubparagraph

\usepackage{subcaption}

\newcommand{\carnaval}{\texttt{CaRNAval}\xspace}
\newcommand{\rnathreedmotif}{\texttt{Rna3Dmotif}\xspace}

\newcommand{\atlas}{\texttt{RNA 3D Motif Atlas}\xspace}

\newcommand{\mb}{\mathbf}

\newcommand{\rthreem}{\texttt{rna3dmotif}\xspace}
\newcommand{\carna}{\texttt{CaRNAval}\xspace}
\newcommand{\vern}{Ve{\scriptsize RNA}l\xspace}
\newcommand{\absvern}{Ve{\scriptsize RNA}l}

\newcommand{\fix}[1]{#1}

\hypersetup{
	colorlinks, linkcolor={red!80!black},
	citecolor={blue}, urlcolor={red!50!black}
}

\begin{document}
\thispagestyle{empty}
	\title{Ve{\normalsize RNA}l: Mining RNA Structures for Fuzzy Base Pairing Network Motifs}
	\titlerunning{\vern}
	
	\author{Carlos Oliver$^\ast$\inst{1,2} \and
		Vincent Mallet$^\ast$ \inst{3,4} \and
		Pericles Philippopoulos\inst{5} \and
		William L. Hamilton\inst{1,2} \and
		J\'er\^ome Waldisp\"uhl\inst{1}
	}
	
	%
	\authorrunning{C. Oliver et al.}
	%
	\institute{School of Computer Science, McGill University, Montreal, Canada \and
		Montreal Institute for Learning Algorithms (MILA) \and
		Structural Bioinformatics Unit, Department of Structural Biology and Chemistry, Institut Pasteur, CNRS UMR3528, C3BI, USR3756 \and
		 Mines ParisTech, Paris-Sciences-et-Lettres Research University, Center for Computational Biology \and
		 Department of Physics, McGill University
		}
	\maketitle              
	\begin{abstract}
    RNA 3D motifs are recurrent substructures, modeled as networks of base pair interactions, which are crucial for understanding structure-function relationships.
    The task of automatically identifying such motifs is computationally hard, and remains a key challenge in the field of RNA structural biology and network analysis.
    State of the art methods solve special cases of the motif problem by constraining the structural variability in occurrences of a motif, and narrowing the substructure search space.\\
    Here, we relax these constraints by posing the motif finding problem as a graph representation learning and clustering task.
    This framing takes advantage of the continuous nature of graph representations to model the flexibility and variability of RNA motifs in an efficient manner.
    We propose a set of node similarity functions, clustering methods, and motif construction algorithms to recover flexible RNA motifs. 
    Our tool, ~\absvern\ can be easily customized by users to desired levels of motif flexibility, abundance and size.
    We show that \absvern\ is able to retrieve and expand known classes of motifs, as well as to propose novel motifs. \\
    \textbf{Availability and Implementation :} The source code, data and a webserver are available at \url{vernal.cs.mcgill.ca}\\
    \textbf{Contact :} jeromew@cs.mgcill.ca\\
	\end{abstract}
	\let\thefootnote\relax\footnotetext{$^\ast$ Both authors contributed equally}
	

\pagenumbering{arabic}

\section{Introduction}

Comparisons of RNA structures revealed the occurrence of highly similar 3D sub-units, called RNA 3D motifs, which are thought to form a basis for non-coding RNA function ~\cite{leontis2006building}.
These motifs are typically characterized by sets of similar base pairing patterns, which are repeated across unrelated RNA \cite{lescoute2005recurrent}.
A complete library of RNA 3D motifs is thus a valuable source of information for evolutionary studies, discovering functional sites, and is an important component of structure prediction methods \cite{sarrazin2019automated,roll2016jar3d}.
Moreover recent advances in small molecule graph generation with deep learning demonstrate the importance of motifs as building blocks \cite{jin2020hierarchical}.
Efficient and automated methods to mine motifs from databases of RNA structures are essential to achieve this goal \cite{djelloul2009algorithmes,reinharz2018mining,petrov2013automated}.

From a methodological point of view, RNA motif mining methods can be placed in two categories: 3D-based and graph-based.
3D-based tools seek to identify families of related structures by performing alignments and clustering of {\it atomic coordinates}.
DARTS \cite{abraham2008analysis},
\atlas \cite{petrov2013automated}, RNA Bricks \cite{chojnowski2014rna}, and  RNA MCS \cite{ge2018novo} illustrate this approach.
Since structural proximity is naturally defined in coordinate space, an advantage of these tools is that variability across occurrences of a motif is achieved for free.
However, these methods require a decomposition of RNA into rigid sub-units to be compared to each other (e.g., comparing all internal loops to each other), which limits the scope of possible motifs to be found.

Alternatively, graph-based approaches work on discrete encodings of RNA 3D structures and rely on network analysis algorithms to extract motifs.
The building blocks of such encodings are linear polymers of nucleotides (A, U, C, G) bound by backbone interactions. These chains determine first, the  highly stable canonical (Watson-Crick and Wobble) and then non-canonical (all other) base base pairing geometries \cite{leontis2001geometric}.
These pairs serve as a scaffold for the formation of the full tertiary structure. 
The conservation of these base pairs is essential to preserving the folding properties of the RNA and offers a robust signature for the functional classification of RNAs \cite{griffiths2003rfam,rnamigos,zhang2016deep}.
Using these components, any RNA 3D structure (set of atomic coordinates) can be represented as a multi-relational graph (also referred to as base pairing network), where nodes correspond to nucleotides, and edges to interactions between nucleotides.
Edge types are assigned based on the classification developed by Westhof et al ~\cite{leontis2001geometric} which defined 12 categories of relative base pair orientations.
These classes can be determined by noting the relative angles and interacting atoms of the bases involved in the pair in 3D space.
In this set of 12 geometries, we can find the standard Watson-Crick (A-U, C-G, G-U) pairs, also known as \say{canonical base pairs}, which are the most stable and abundant class.
However, when interpreting 3D motifs, the remaining 11 geometries, also known as \say{non canonical} are necessary for understanding RNA structure at a 3D level \cite{leontis2001geometric}.

\fix{Graph-based tools therefore typically aim to identify similarities at the base pairing level.
Of course, identifying motifs requires a combinatorial search and hence such tools impose strong limitations on the search over subgraphs.
Chief among these is the ability to include variability within motifs.
The notion of a fuzzy motif has been very well studied in the sequence domain \cite{d2006dna} where certain DNA sequences are accepted to be related while their nucleotide composition can vary.
Not surprisingly, the same applies in the RNA structural domain where well-studied motifs such as the A-minor are known to admit variability in their connectivity pattern \cite{aminor}
Closely related 3D structures may be represented as quasi isomorphic (\fix{or }fuzzy) graphs.
However, most classic and tractable graph comparison algorithm rely on exact matching.
Methods which rely on strict isomorphism would fail to identify fuzzy instances, as well as fail to discover some motifs entirely.
Another limitation here is that evaluating all potential graph motifs involves searching over the set of all subgraphs which grows exponentially in the graph size.
A first approach to resolve this challenge is to use already known motifs as a queries to search for new instances (RMDetect \cite{rmdetect}, RNAMotifscan \cite{zhong2010rnamotifscan,zhong2015rnamotifscanx,blast3d,frabase}).
However these methods require motifs as input, which are typically obtained through visual inspection such as A-minor or kink-turn motifs.
There is thus a need for {\it de-novo} motif mining tools that look into the space of subgraphs and find the recurrent ones.
To avoid enumerating all subgraphs, many motif mining tools restrict themselves to specific substructures.
A first reach for {\it de-novo} motifs is thus conducted by \cite{lemieux2006automated} working only on a single ribosomal unit, and focusing on cycle motifs, a very specific and small subgraph type.
\rthreem \cite{djelloul2009algorithmes} offered the first library of exact motifs only within certain known structural elements, namely the k-way junction.
Another approach to this problem is metaRNAmodules \cite{metarnamodules} which enumerates all nested loops and uses RMDetect with a statistical to filter the recurrent ones.
Another graph-based method, developed in 2015 is RAG-3D ~\cite{zahran2015rag} which uses a graph abstraction to mine motifs spanning multiple secondary structure elements, and simultaneously proposes a query-search functionality a discussed above.
More recently, \carna \cite{reinharz2018mining,soule2021finding} attempted to expand the class of motifs by considering interactions that connect multiple secondary structure elements while maintaining isomorphic motif instances. 
All these tools either focus on local motifs or impose a strict isomorphism of motif occurrences.
}

\subsection*{Contributions}

We leverage the state of the art in graph representation learning to build continuous embeddings of RNA substructures and identify structurally conserved yet variable motifs.
We then propose two algorithms that use these graph representations to find graphs similar to a query and to identify novel motifs.
\fix{By comparing with existing motif libraries, we are able to efficiently identify unknown instances of existing motifs, and propose over 1,800 densely populated motifs for further exploration.}

\section{Datasets}

We extract motifs from the set of experimentally determined RNA structures \cite{rose2016rcsb}.
To ensure that the frequency of a motif is not biased by redundant  structures, we use the representative set at 4 Angstrom resolution provided by the BGSU RNA 3D Hub \cite{petrov2013automated}.
We then build an RNA network with 13 edge types for each RNA using the FR3D annotations provided by the same framework.
Each edge represents either a backbone (covalent) interaction, or a base pair classified in 12 geometries according to the relative orientation of the interacting bases (nodes).
This results in a total of 899 RNA graphs and 210616 nodes (nucleotides).
In the learning phase, we chop these graphs in constant-sized chunks of approximately 50 nucleotides to avoid dealing with graphs of heterogeneous sizes, as is detailed in \textbf{Supplementary Algorithm \ref{supp:algo:chopper}}.
Once the model is trained we perform all motif finding operations on whole graphs.
Our validation sets consist of motifs identified by \atlas  \cite{petrov2013automated}, \rthreem \cite{djelloul2009algorithmes}, and \carna \cite{reinharz2018mining}.

\section{Methods}

We introduce \vern, an algorithm to efficiently identify fuzzy recurrent network motifs in RNA.
\vern decomposes RNA networks into structural building blocks and then aggregates these blocks based on their co-occurrence in RNA. 

The decomposition step introduces custom structural comparison functions which are used to build a space of continuous embeddings for efficient clustering (Section \ref{sect:embs}). 
We then combine information from the embedding space and connectivity in the graph space into a meta-graph data structure (Section \ref{sect:meta-graph}). 
We leverage this data structure to retrieve graphs similar to a query (Section \ref{sect:retrieve}), and to streamline frequent substructure searches and thus identify \textit{fuzzy} motifs (Section \ref{sect:maga}).

\subsection{Problem Definition}


As described above, each RNA 3D structure can be encoded into a multi relational graph.
Without loss of generality, we consider that the set of all such graphs forms one large disconnected directed graph $\mathbb{G} = (\mathbb{V}, \mathbb{E})$ with about 670k nodes.
We define a motif as a set of subgraphs $\mathcal{M} = \{g_1, g_2,..\}$, drawn from $\mathbb{G}$, such that the following properties hold:

\begin{enumerate}
    \item {\it Similar:} Let $s$ be a similarity function on graphs, and $\gamma \in [0,1]$. $ \forall (g_i, g_j) \in \mathcal{M}^2$, $s(g_i, g_j) \geq \gamma$
    \item {\it Connected:}  $\forall g_i \in \mathcal{M}$, $g_i$ is a connected subgraph.
    \item {\it Frequent:} the number of subgraphs of $\mathcal{M}$ should be above some user-defined threshold : $| \mathcal{M} | > \delta$
\end{enumerate}

The motif mining problem is simply to identify all sets of subgraphs (motifs) $\mathcal{M} \in \mathbb{G}$ that fit the above criteria.
An exact solution to this problem would imply enumerating all subsets (search for subgraphs) of $\mathbb{G}$ and ensuring that these criteria are satisfied (compare graphs).
In the most general case, both procedures admit exponential time algorithms \cite{Zeng2009ComparingSO}.
Previous works set $\gamma = 1$,  so that the similarity constraint becomes another graph problem, known as the subgraph isomorphism problem ~\cite{reinharz2018mining,djelloul2009algorithmes}. 
This also constrains these methods to rely only on pairwise comparisons and prevents them to detect communities of close but different neighbors.
Additionally, the search step is often limited by considering only certain substructures (e.g. hairpins, internal loops, etc).
We can consider two additional properties of motifs: \textit{maximality}, and size.
A motif is said to be {\it maximal} if adding a node to $g_i$ breaks the other motif constraints.
We define the size of a motif as the mean number of nodes per graph in $\mathcal{M}$. 
Enforcing these two properties is left as an implementation choice.
For example \carnaval ~\cite{reinharz2018mining} returns maximal exact subgraphs containing long range interactions with no size constraints.
Here, we remove search constraints on secondary structure context ~\cite{petrov2013automated,reinharz2018mining}.
We also allow for non-identity $\gamma$ and call this property {\it fuzziness}. 

\begin{figure*}[t!]
    \centering
    \includegraphics[width=\textwidth]{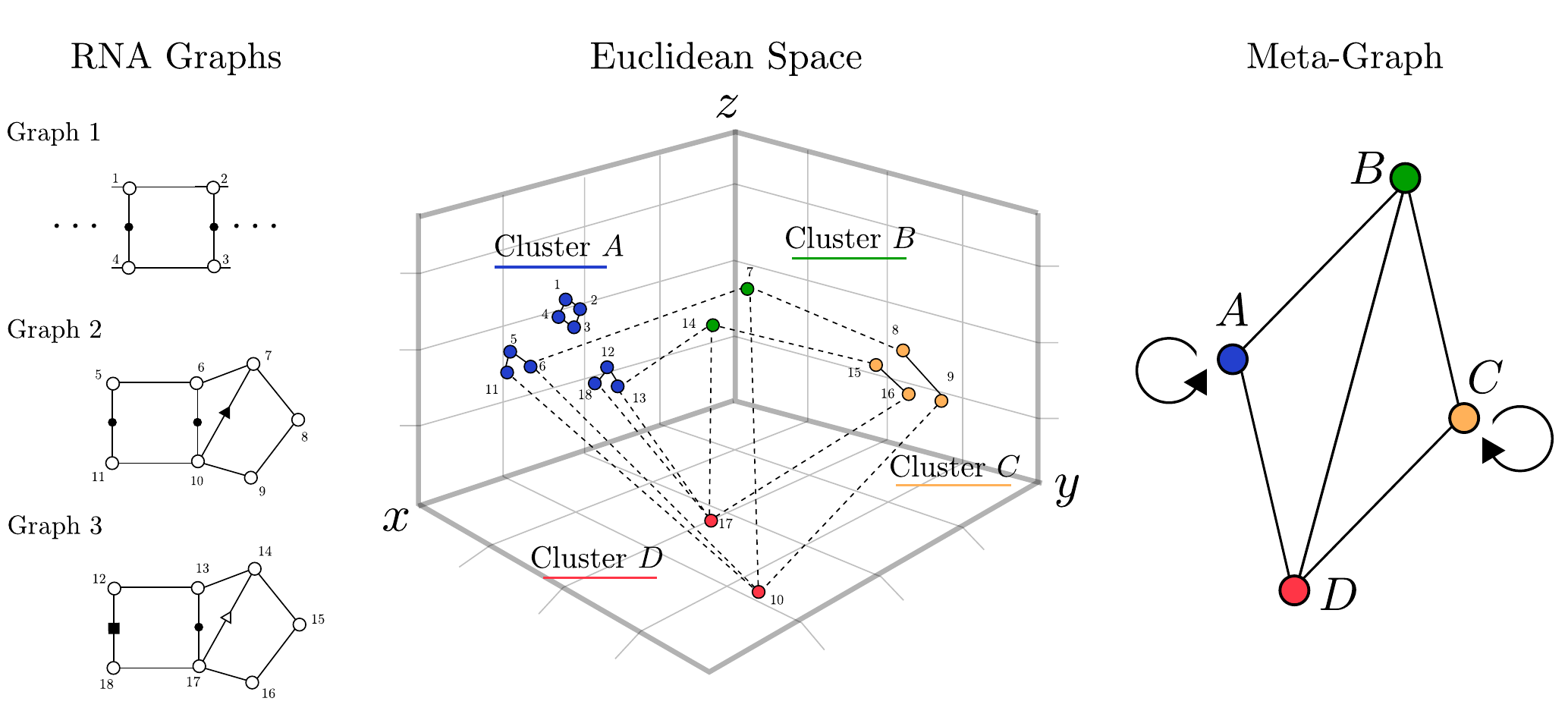}
    \caption{Meta-graph creation : RNA graphs \textit{(Left)} get aligned in the embeddings space \textit{(Middle)} and represented as a meta-graph \textit{(Right)}. RNA nodes are grouped in meta-nodes through clustering, which reflects structural similarity. Meta-edges are then inferred from the source graphs connectivity}
    \label{fig:meta_graph_sketch}
\end{figure*}

\subsection{Rooted Subgraph Embeddings}
\label{sect:embs}
Maximal subgraph isomorphism algorithms rely on heuristics that are not applicable anymore for $\gamma <1$.
Allowing fuzziness calls for other efficient ways to compare graphs.
We turn to recent advances in Graph Representation Learning, which provide a framework for encoding rooted subgraphs ~\cite{hamilton2017representation}.
A rooted subgraph $g_u$ is the induced subgraph on the set of nodes $u' \in g$ such that $p(u, u') \leq r$ where $p$ is the length of the shortest path between two nodes, and $r$ is a user-defined threshold (also known as radius).
A vector embedding of dimension $d$ for a given $g_u$ is computed by a parametric function $\phi : g_u \rightarrow \mathbb{R}^d$ (typically a Graph Neural Network).
These embeddings seek to approximate a graph similarity function $s_G: g_u \times g_v \rightarrow [0, 1]$ acting directly on graphs.
In an unsupervised setting, $\phi$ is trained to minimize the loss described in Equation \ref{eq:loss:main}.
The resulting embedding space conveniently captures the desired property of fuzziness, as proximity in the embedding space (via inner product) corresponds to structurally similar nodes in the graph space.
\begin{equation}
    \mathcal{L} = \| \langle \phi(g_u), \phi(g_{u'}) \rangle - s_G(g_u, g_{u'}) \|^2_2,
     \label{eq:loss:main}
\end{equation}

For RNA motifs, we are only interested in considering edge type and graph structure.
Notably, it is known that certain base pairing geometries (edge types) share structural similarities : a phenomenon known as isostericity (Supplementary Section \ref{supp:sect:isostericity}).
We introduce various customized similarity functions on RNA graphs which account for key 3D geometric features such as isostericity \cite{stombaugh2009frequency} and base pairing type \cite{leontis2001geometric}.
They are based on the matching of sub-parts of the rooted subgraphs as computed using the Hungarian algorithm \cite{Kuhn55thehungarian}.
We then use a Relational Graph Convolutional Network (RGCN) model \cite{schlichtkrull2018modeling} as the parametric mapping.
The network is implemented in Pytorch \cite{paszke2017automatic} and DGL \cite{wang2019deep}. 
To focus performance on subgraphs that contain non canonical nodes and avoid the loss to be flooded by the canonical interactions (Watson-Crick pairs), we then scale this loss based on the presence of non canonical interactions in the neighborhood of each node being compared. 
Details on the similarity functions and the training of the network are included in the Supplementary Sections \ref{supp:sect:kernels} and \ref{supp:sect:rgcn}.
We can then perform clustering in the embedding space\fix{, using the k-means algorithm \cite{macqueen1967some}.
We select the number of clusters according to the Silhouette Score and several clustering metrics (See \textbf{Figure \ref{supp:fig:clustering}})}.
We denote the resulting clusters as 1-motifs (rooted subgraphs with 1 root) as they represent the aforementioned structural blocks of RNA.

\subsection{Meta-graph}
\label{sect:meta-graph}

While there is no limit to the size of a real-world motif, our rooted subgraph embeddings are currently only aware of a fixed-size neighborhood, (i.e. the radius of the rooted subgraphs). 
For this reason, these clusters only identify motifs as large as the \fix{radius of the rooted graphs}. 
They are centered around just one node so we denote them as 1-motifs.
However, we want to extend these to $k$-motifs by aggregating $k$ different nodes based on co-occurrence in the original graph. 

To guide this aggregation, we introduce a meta-graph data structure $\mathcal{G}$ \fix{$= (\mathcal{C}, \mathcal{E})$}, whose meta-nodes are composed of regions of the embedding space and whose meta-edges are based on the connectivity in the RNA graphs between those regions.
\fix{Meta edges are weighted by the amount of existing edges they represent.}
Hence, the meta-graph simultaneously encodes structural proximity and connectivity in the graph in one object.
We can see it as a coarsened version of $\mathbb{G}$ when its nodes are structurally embedded in \fix{the Euclidean space}.
If two meta-nodes are connected by a large meta-edge, it means that a these two structural elements are often adjacent in the graph and thus, they are good candidates to be merged.

To get the meta-nodes, we embed the original nodes in $\mathbb{V}$ into $\mathbb{R}^d$. 
We then cluster these embeddings and use the clusters as meta-nodes : $C_i = \{u\in\mathbb{V},\ \text{cluster}(\fix{\phi(g_u)})=i\}$, where $C_i$ is the $i-th$ node in $\mathcal{G}$. 
Choosing the number of clusters and their spread modulates the fuzziness of the resulting motifs.
Meta-edges $E_{i,j} = \{ (u_i, u_j) \in (C_i \times C_j) \cap \mathbb{E} \} $ store the edges in RNA graphs that go from one cluster to another.
This process is illustrated in \textbf{Figure \ref{fig:meta_graph_sketch}}.
Building the meta-graph requires RGCN inference and clustering over $\mathbb{V}$, and iterating through all edges in $\mathbb{G}$.
With linear-time clustering techniques (like k-Means \cite{macqueen1967some} or Gaussian Mixture), building the meta-graph is therefore done in time $O(|\mathbb{V}| + |\mathbb{E}|)$.

This meta-graph data structure helps us merging clusters together based on their connectivity.
Merging meta-nodes $C_1$ and $C_2$ results in a 2-meta-node which contains all subgraphs in $C_1$ that are connected to a subgraph in $C_2$.
These subgraphs are indexed in the meta-edge $E_{1,2}$.
However, inferring the connectivity of the merged meta-node is not trivial.
\fix{Indeed, the neighbors of a merged meta-node are not the union of the neighbors of its constituent meta-nodes, since we retain only a subset of the graphs of each merged meta-node.}
This problem is illustrated in \textbf{Figure \ref{fig:maga_sketch}} where the merged meta-node BC is not connected to A despite A and B being connected.
To address this problem, \fix{we only consider adding 1-meta-nodes to an arbitrary meta-node. 
This is not a limitation for our algorithm and is more compatible with our formalism}
We implement a merging algorithm presented in \textbf{Algorithm \ref{algo:merging}}.

\subsection{Retrieving known motifs}
\label{sect:retrieve}
The first use of the meta-graph data structure is to retrieve subgraphs similar to a query subgraph.
Given a query subgraph $Q$ and a large disconnected graph $\mathbb{G}$, the task is to identify all hits : subgraphs $H \in \mathbb{G}$ which maximize similarity to the query.
The idea of the algorithm is to use the 'alignment' of the RNA graphs induced by the embeddings ({\bf Figure ~\ref{fig:meta_graph_sketch}}) to efficiently search for similar structures.
Such an algorithm can identify subgraphs that resemble known motifs but which were not identified by tools imposing strict isomorphism \cite{rmdetect,zhong2010rnamotifscan,zhong2015rnamotifscanx}. 

Using the RGCN, we place the query graph in the embedding space, which induces a query meta-graph $\mathcal{Q}$.
\fix{This meta-graph is a subgraph of $\mathcal{G}$, each query meta-node is defined by the cluster assignment of the embedding of each nucleotide  of the query graph.
Then for each edge in the query graph, we add a meta-edge between its corresponding meta-nodes.}
Let $c_i$ be the centroid of a meta-node $C_i$.
We can directly obtain a compatibility score between a query meta-node $q \in \mathcal{Q}$ and a hit node $h \in \mathbb{G}$ : $score =  \langle \phi(g_h), c_q \rangle$.
We start with all nodes in a meta-node of $\mathcal{Q}$, as one-node hits.
The one-node hits and their scores are added to a set $\fix{\mathcal{R}}$.

To expand the match, we iterate through the edges of the $\mathcal{Q}$ and merge any two elements of $\fix{\mathcal{R}}$ that fall along the current edge.
We do so by using the aforementioned merging algorithm.
We can think of this step as iteratively building bigger and bigger sub-graphs of $\mathbb{G}$ that match the template provided by $\mathcal{Q}$.
Any merge operation increases the score of the resulting set by summing the score of the merged elements.
If a hit encompasses all nodes in the query, it will have undergone the most merging operations and obtain a maximal score.
However, if a hit misses one node or has a somewhat different structure, we still retrieve it with a sub-optimal but high score.
This retrieval procedure is detailed in \textbf{Algorithm \ref{algo:retrieval}}.

\begin{figure}
\begin{minipage}[t]{.45\textwidth}
\begingroup
\makeatletter
\let\@latex@error\@gobble
\makeatother
\begin{algorithm}[H]
\SetAlgoLined
\KwData{\begin{itemize}
	\item Meta-node S
	\item Meta-edge $E$ from $S$ to $C$
\end{itemize}
}
\KwResult{$T$, an expanded $S$ along $E$}
$T \leftarrow \varnothing$ \\
\ForEach {$(v_1,v_2) \in E $ }{
	\ForEach {$nodeset \in S $}{
    	\If{$ \{v_1, v_2\} \Delta nodeset \neq \varnothing $\footnote{$\Delta$ represents the symmetric difference of two sets. Here we mean if precisely one of the nodes is already in the nodeset.}}{
    	$T \leftarrow T \cup \{ nodeset \cup \{v_1\} \cup \{v_2\} \}$ 
    	}
    }
}
\Return $T$
\caption{Merging Algorithm.
We merge a meta-node $S$ with a 1-meta-node $C$, through a meta-edge $E$.
}
\label{algo:merging}
\end{algorithm}
\endgroup
\end{minipage}%
\hfill
\begin{minipage}[t]{.5\textwidth}
\begingroup
\makeatletter
\let\@latex@error\@gobble
\makeatother
\vspace{0pt}
\begin{algorithm}[H]
\SetAlgoLined
\KwData{\begin{itemize}
	\item Meta-graph $\mathcal{G} = (\mathcal{C},\mathcal{E})$
	\item Original RNA graphs $\mathbb{G} = (\mathbb{V}, \mathbb{E})$
    \item Query multi-graph $\mathcal{Q}$
\end{itemize}
}
\KwResult{$\fix{\mathcal{R}}$ : Motif instances candidates : a set of subgraphs and their associated scores} 
$\fix{\mathcal{R}} \leftarrow \bigcup\limits_{C \in \mathcal{Q}} C$ \\
\ForEach{E in $\mathcal{Q}$}{
	$T \leftarrow \texttt{merge}(\fix{\mathcal{R}}, E$)\\
	$\fix{\mathcal{R}} \leftarrow \fix{\mathcal{R}} \cup T$\\
}
\Return $\fix{\mathcal{R}}$
\caption{Motif Instances Retrieval. We traverse the edges of the graph $\mathcal{Q}$ and the meta-graph identifying connected subgraphs which match query embeddings.}
\label{algo:retrieval}
\end{algorithm}
\endgroup
\end{minipage}
\end{figure}

The algorithm remains tractable thanks to the sparsity of the meta-graph that allows efficient iteration through edges, efficient set operations to expand motifs and graph-based separation of the candidate hits.
A theoretical analysis of the complexity depends heavily on both the topology of the meta graph and of the query graph and is explained further in Supplementary Section \ref{supp:sect:complexity}. 
We observe that in practice, this algorithm runs in an average of 10s on a single i7-10610U core.

\subsection{Mining new motifs}
\label{sect:maga}

We can leverage a similar strategy to the retrieve procedure when mining motifs {\it de novo}.
The basic intuition of our algorithm, Motif Aggregation Algorithm (MAA) is again that the set of nodes assigned to a given cluster can be considered to be a motif of cardinality 1 (a 1-motif).
We can then use the meta-graph to identify clusters with connections to the current motif set to build larger motifs.
Because we lack the guidance of the query, instead of merging just along one meta-edge, we merge along all meta-edges in the meta-graph and filter results based on a user-defined minimal frequency $\delta$.

\begin{figure}
    \centering
    \includegraphics[width=0.5\textwidth]{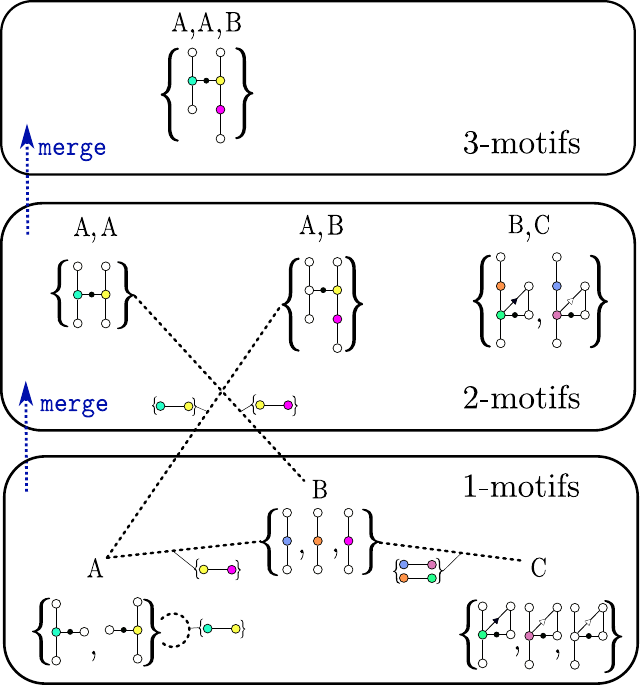}
    \caption{MAA Illustration : Meta-nodes A,B and C get merged into three 2-Meta-nodes AA, AB and BC. Then new meta-edges are computed that link singletons A and B with 2-meta-nodes AB and AA  respectively. A second merge follows these links and yields the 3-meta-node AAB. Node colors here are a proxy for node ID, and not tied to the cluster IDs (A, B, C).}
    \label{fig:maga_sketch}
\end{figure}

As an example, starting with a 1-motif e.g. the set of subgraphs in cluster $A$, we can create 2-motifs by merging every other cluster in its meta-graph neighborhood, $X \in \mathcal{N}(A)$. 
We then identify the new 2-motifs from their constituent meta-nodes.
This process can then be iterated to discover $k$-motifs.
This is illustrated in \textbf{Figure \ref{fig:maga_sketch}} and outlined in detail in \textbf{Supplementary Algorithm \ref{algo:motif}}.

Given the  state of the meta-graph $\mathcal{G}$ at step $k$, identifying all instances of a single $(k+1)$ motif requires a single \texttt{merge} operation between two meta nodes, with complexity $\mathcal{O}(n)$.
The number of all such possible $(k+1)$ motifs in the worst case is $\mathcal{O}\big(\binom{\fix{|\mathcal{C}|}}{k}\big)$ which is the number of choices of $(k+1)$-size meta nodes.
However, this is a loose bound since in practice, the number of acceptable motifs is constrained by the sparsity of $\mathcal{G}$ at $k=0$ and the minimum motif frequency $\delta$.
Empirical complexity depends strongly on hyper-parameters choices but is on average a few minutes on a single i7-10610U core.

\section{Results}

Our tool relies on graph representation methods to drastically improve the scalability of motif mining and facilitate fuzzy matching of motifs. 
We first evaluate the quality of our RNA-specific similarity functions and subsequent embeddings (Section \ref{sect:validation_embs}) and show that structural information is faithfully encoded.
Following this, we show that our approach can consistently retrieve existing motifs (Section \ref{sect:validation_retrieve}) while also uncovering new fuzzy motifs (Section \ref{sect:validation_maa}).
Throughout the evaluation of the tools, we use Graph Edit Distance (GED) ~\cite{bunke2008graph} as an external (and costly) oracle to select a similarity function, assess embedding quality, and motif consistency.
For more details on GED definition and implementation see Supplementary Section \ref{supp:sec:ged}.

\subsection{Subgraph Comparisons and Embeddings Correlate with GED}

\label{sect:validation_embs}

We sample 200 rooted subgraphs of radius 1 and 2 uniformly at random from $\mathbb{G}$.
We recall that the radius of a graph is the maximum length shortest path between any two nodes in this graph.
Next, we compute all-to-all GED on this sample, yielding 20,000 non trivial values for each radius.
We then compute similarities on the same set of subgraphs using various choices of $s_G$ and $\phi$.
In {\bf Supplementary Table ~\ref{table:main_ged_corr}} we summarize the resulting Pearson correlation values.

Under these metrics, the best performing method performs a matching over ordered sets of smaller graphs known as graphlets, for more details see Supplementary Section \ref{supp:sect:kernels}.
It gets an almost perfect correlation at a radius of one and 0.52 at a radius of two. 
Since we consider fuzzy motifs to consist of graphs with slight variations, performance on similar graphs is more relevant.
On pairs of graphs with low GED to each other, we obtain higher correlations of 0.637 on the radius-two subgraphs.
Next, we train a 2-layers RGCN using this similarity function and obtain a thresholded correlation value of 0.74 with the GED values. 
Therefore, the dot product of our embeddings approximates structural similarities.
Moreover, we note that the running time of a comparison becomes negligible, as it amounts to a dot product.
To simplify downstream analysis, we take advantage of the strong correlation and use only 1-hop rooted subgraphs.
Full results are available in \textbf{Supplementary Table \ref{table:main_ged_corr}}.

\begin{figure}
	\centering
	\includegraphics[width=.7\textwidth]{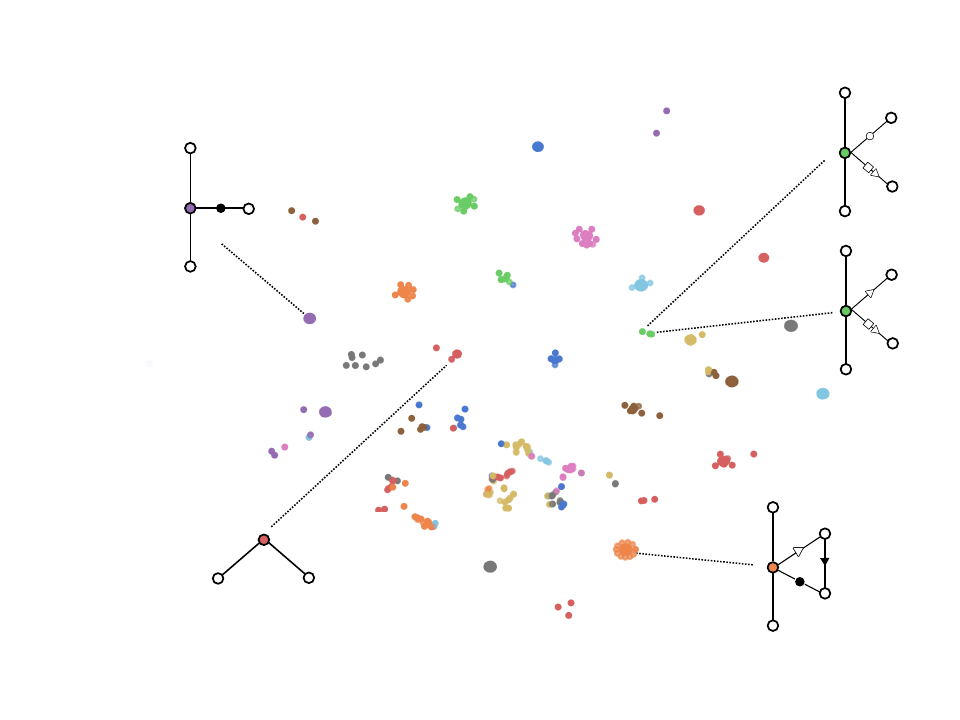}
	\caption{t-SNE projection applied to embedding space. Drawn rooted subgraphs correspond to an example from the cluster connected by a dotted line. Point colors correspond to the nearest mixture model component.}
	\label{fig:space}
\end{figure}


We complete our report of the performance assessment of the embeddings with a visual representation of the results.
In \textbf{Figure \ref{fig:space}}, we generate a 2D projection of the local RNA structures from the learned embedding with t-SNE \cite{tsne}.
 We draw example subgraphs corresponding to a sample of clusters.

Visually, we observe that similar subgraphs lie in the same clusters. 
Additional quantitative metrics are provided in the Supplementary Section \ref{supp:sect:clusts}. 
This validation provides us the structural building blocks to assemble and retrieve motifs.

\subsection{Retrieval Algorithm Expands Known Motifs}
\label{sect:validation_retrieve}

Next, we turn to the validation of the retrieval algorithm. 
Given a query graph, the retrieve algorithm returns a list of subgraphs of $\mathbb{G}$, denoted as \say{hits}, in decreasing order of compatibility to the query.
We run the algorithm \fix{on our validation set of motifs (motifs identified by \atlas  \cite{petrov2013automated}, \rthreem \cite{djelloul2009algorithmes}, and \carna \cite{reinharz2018mining}), filtered for sparsity (more than 3 instances) and size (more than 4 nodes), resulting in 285 motifs.} 
For a given known motif, we perform a retrieve with two types of queries: a true instance of the motif, and an instance of another randomly chosen motif (decoy).
We show the resulting ranks in \textbf{Table \ref{table:retrieve1}}.
The hit list contains a few good hits and a long tail of very small hits.
We see that when queried with a true instance, the algorithm retrieves other instances in \fix{97\%} of the cases, \fix{with the average rank in the top 5\%}. 
When queried with a decoy, the success rate drops to 83\%, with an average rank at the 24th percentile of the hit list, indicating that only partial solutions were retrieved.

We can go further by analyzing the structure of the retrieved hits.
A first way to do so is to plot several hits with increasing ranks (\textbf{Figure \ref{fig:smooth_retrieve}}). 
A visual inspection of the results indicates that the retrieved graphs differ more and more as we plot hits with decreasing scores. 
A more quantitative way to do this is to compute the mean GED value of hits at fixed ranks compared to their respective queries.
\fix{Since the motifs contain up to 15 nodes, the GED computation is not always exact and can yield high running times with incorrectly high values. 
We discard motifs that reached the GED timeout and end up with a set of 140 motifs, for which we present the mean GED with their hits in \textbf{Table \ref{table:retrieve2}}.
For comparison, we also include the GED to a decoy corrected to match the size of the query.
Sometimes, the best hits are not exactly isomorphic but happen in a more similar context than other isomorphic graphs, resulting in closer embeddings for the nodes at the border of the motif.
This happens more frequently for larger graphs and explains why the GED is not a hard zero, but we see that it is very significantly shifted towards lower values.}
Based on both of these results we claim that our method is able to retrieve sets of subgraphs where the GED to the query correlates with the retrieval rank.

\begin{table}
    \begin{minipage}{.5\linewidth}
        \begin{center}
            
\begin{tabular}{lrr}
\toprule
         Method &  Success Rate &  Normalized rank \\
\midrule
 True query  &  97\% & 5,6\%  \\
 Decoy query &  83\% & 23\%  \\
\bottomrule
\end{tabular}
\caption{Comparison of the performance of the retrieve algorithm when used with a query instance vs. a random one. Success rate denotes the rate at which the instance is in the hit list. The rank denotes the rank in the list, normalized by its length (lower is better). }
\label{table:retrieve1}

        \end{center}
        \begin{center}
            
\begin{tabular}{lrrrrr}
\toprule
Rank &  $1^{st}$ &  $10^{th}$ &  $100^{th}$ & $1000^{th}$ & Decoy\\
\midrule
Mean GED &   $3.1\pm 0.3$ &  $3.9\pm 0.4$ &  $6.2\pm 0.6$ & $9.2\pm 0.8$ & 14.4 $\pm 0.8$ \\
\bottomrule
\end{tabular}
\caption{Mean GED with standard errors between motifs queries and their hits at fixed ranks. We also included mean GED values to other random motifs as a control.}
\label{table:retrieve2}

        \end{center}
    \end{minipage}\hfill
	\begin{minipage}{0.45\linewidth}
		\centering
        \includegraphics[width=\linewidth]{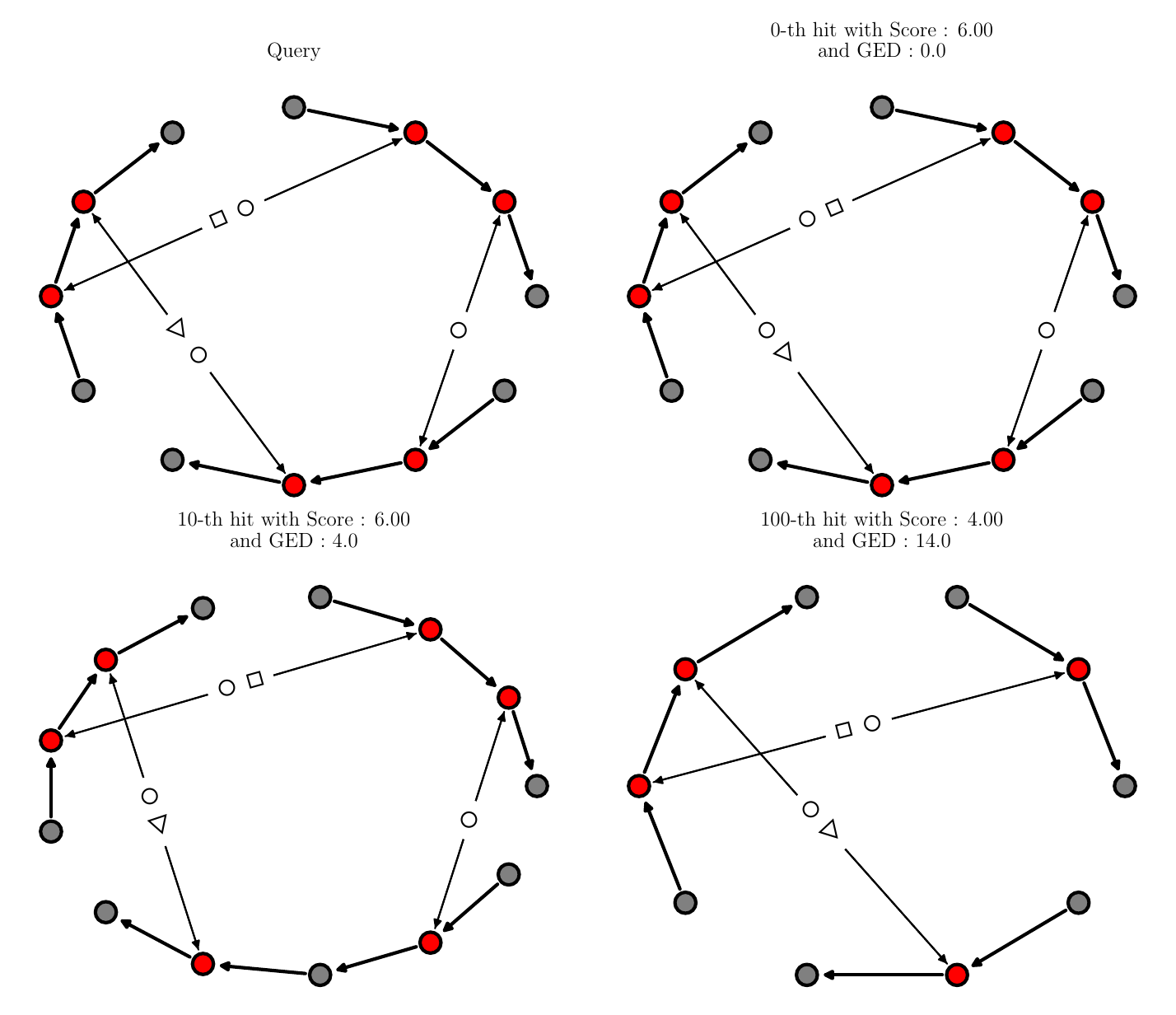}
		\captionof{figure}{Hit graphs with decreasing rank to the query. Red nodes indicate matches to the query.}
		\label{fig:smooth_retrieve}
	\end{minipage}
\end{table}


The average number of instances of a motif across \rthreem, \atlas, and \carna\  is only 22.3.
Interestingly, the fact that we we are able to obtain up to 100 hits with a low GED indicates that many of these represent an ensemble of highly similar structures that are missed by existing tools.
This observation suggests that our method can be used not only to assess if we find known instances of a motif, but also to identify fuzzy instances of these well known motifs. 

\subsection{MAA Identifies Novel Fuzzy Motifs}
\label{sect:validation_maa}

Finally, we assess the quality of the MAA procedure to identify {\it de novo} motifs.
Of course, there are many choices of hyperparameters which are ultimately application-dependent (fuzziness, motif frequency, size, etc.).
We require a minimum frequency of 100 instances per motif, as well as a maximum cluster spread of 0.4 in units of Euclidean distance.
Finally, we remove a motif if more than 80\% of its instances are included in a bigger motif, to enforce \textit{maximality} of the retrieved results.
\fix{We obtain a set of 3,496  motifs up to cardinality 7.}
{\bf Supplementary Table \ref{supp:table:sizes}} shows the average number of instances and number of motifs at each cardinality.

\begin{figure*}
    \centering
    \begin{subfigure}[t]{.32\textwidth}
    \centering
    \includegraphics[width=\textwidth]{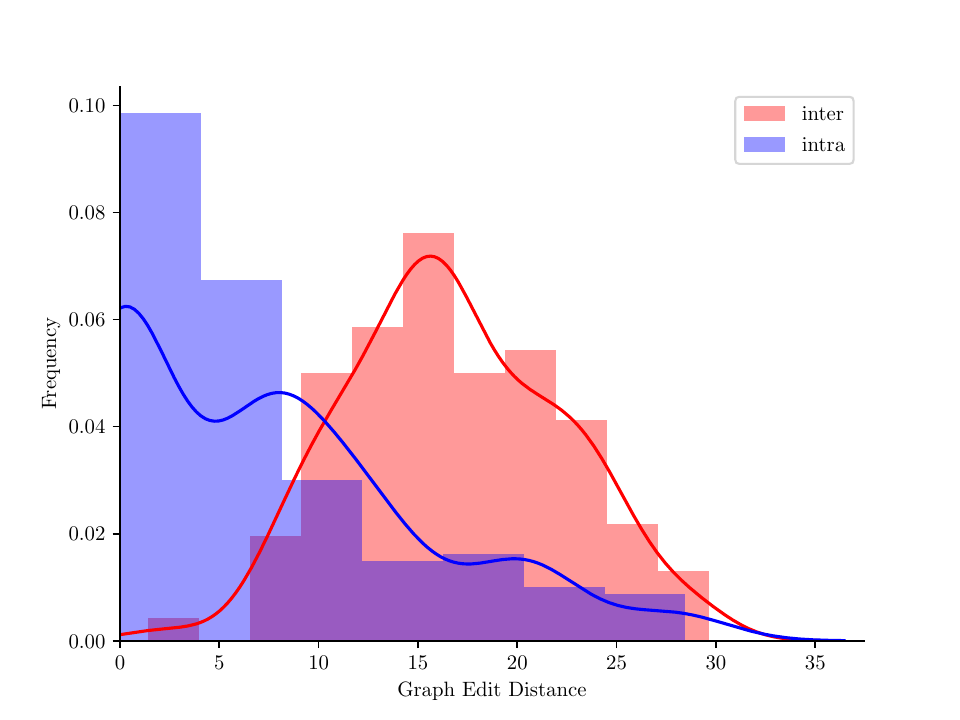}
    \caption{Distributions of GED for subgraphs sampled within the same motif (intra) and across motifs (inter).}
    \label{fig:inter-intra}
    \end{subfigure}
    \hfill
    \begin{subfigure}[t]{.66\textwidth}
    \centering
    \includegraphics[width=\textwidth]{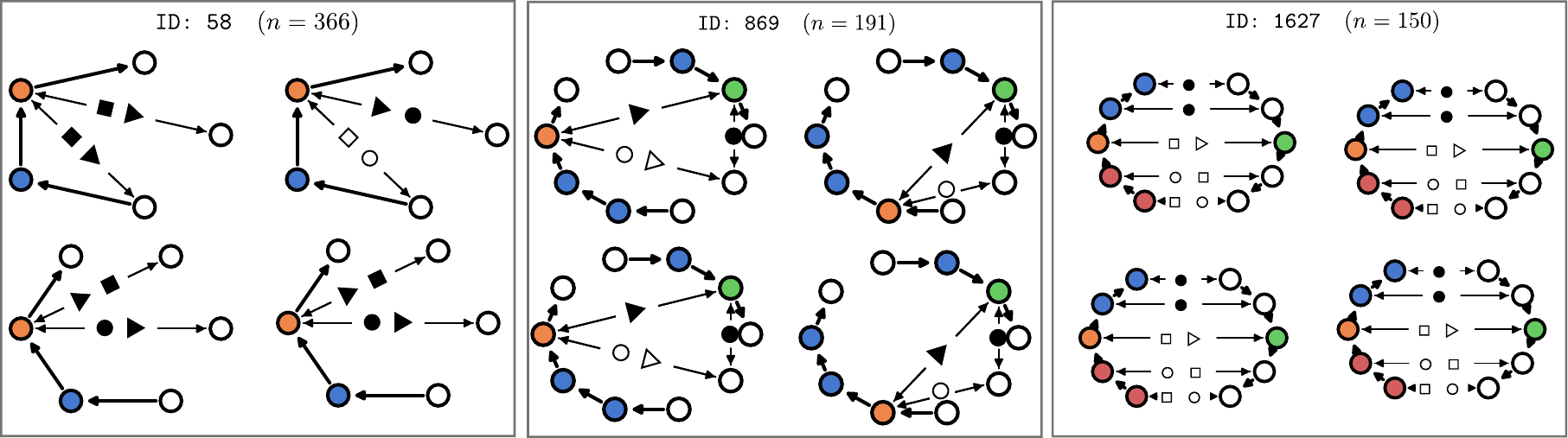}
    \caption{Four instances from three random \vern\ motifs that did not overlap with external motifs. Each root node's color corresponds to its cluster ID.}
    \label{fig:sample-motifs}
    \end{subfigure}
    \caption{\vern motif quality, as measured by GED, and sample novel motifs.}
\end{figure*}

To check for internal consistency, we compute the intra- and inter-motif GED between a random sample of 20 motifs and plot the results in \textbf{Figure \ref{fig:inter-intra}}.
\fix{We obtain an intra-motif GED of 7.27 $\pm$ 7.63 and an inter-motif GED of 16.78 $\pm$ 5.42  when comparing motifs of the same size.}
This shows that  \vern\ finds motifs with internal consistency.

Next, we measure the degree to which our motif set agrees with existing motif databases.
Since our approach is a generalization of RNA motifs (with $\gamma \leq 1$), we expect that known motifs would form a subset of \vern motifs.
Indeed, we find that a subset of our motifs aligns well (at least 60\% overlap in number of nucleotides) with all databases, with a total of 82\% of known motifs being found as \vern motifs (detailed results in \textbf{Supplementary Table \ref{table:hmp}}).
In the heatmap of \textbf{Supplementary Figure \ref{fig:overlap}}, we plot the percentage of nodes of a known motif (\atlas (alias: Atlas), \carna, and \rthreem) that can be found in any of our motifs of the same size.
Noticeably, our tool automatically retrieves motifs previously identified using several different constraints, methods and objectives.  Indeed, \carnaval focuses on motifs with long range interactions, while \rnathreedmotif targets motifs within loops.  
It shows that our framework unifies previous approaches but also expands their reach.
Additionally, the \vern\ motifs that match known motifs feature many more instances, again suggesting that we are able to expand the set of known motif instances. 
Finally, despite no runtime being reported by \carnaval, the authors indicated through personal communications that their analysis ran in about 300 hours. By contrast, \vern used about one hour of training (once) and a couple of minutes for running the clustering algorithm.

\fix{Since our motif set is significantly larger than those presented by existing methods, it is possible that the tool is uncovering novel motifs and exploring new regions of the structure space}. 
A cursory examination of randomly sampled motifs reveals potentially interesting relationships between RNAs of varying functional roles.
\fix{For example, motif \texttt{ID: 869} (drawn in {\bf Figure ~\ref{fig:sample-motifs}})  contains 191 occurrences, mostly from ribosomal RNA, but interestingly some instances fall in a viral ribozyme complex (PDBID: 1Y0Q).
Similarly, motif $\texttt{ID: 1627}$ features instances from ribosomes, as well as the FMN riboswitch (PDBID: 3F2X).}
An in-depth analysis of all individual instances is out of the scope of this contribution, but we plot some additional examples in {\bf Supplementary Figure \ref{fig:supp:sample-motifs}}.
Nonetheless, all the motifs identified by \vern\ can be browsed and downloaded on our web server \url{vernal.cs.mcgill.ca}, and are thus available to the community for further analysis.

\FloatBarrier

\section{Conclusions}

We describe \vern, a novel pipeline for identifying fuzzy graph motifs.
We develop various node structure comparison functions and approximate their feature map using an RGCN, embedding our graph dataset to a vector space for fast similarity computation between rooted subgraphs. 
We show that these computations correlate well with the RNA GED while being significantly faster.
This enables us to find small structural building blocks of RNA and organize them into a meta-graph data structure.

Using this custom data structure, we introduce two algorithms to retrieve similar instances to a known query and to discover new motifs.
We show that the retrieval procedure enables us to efficiently identify other instances of known motifs but also to find sets of subgraphs similar but not identical to a query.
The motif extraction algorithm is also successful in mining sets of subgraphs with low intra-cluster GED, re-discovering and expanding known motifs as well as introducing new ones. All together, our platform \vern\ is the first tool to propose fuzzy graph motif extraction.

The nature of graph convolutions somewhat limits the type of motif that can be detected by \vern.
Since RGCNs perform convolutions of entire neighborhoods around a node, motifs without a wide-enough conserved core can be lost, as information from outside the motif gets aggregated together with frequent nodes.
Additional tuning of the similarity function radius, or more advanced message passing methods could address this limitation.

The main focus of this work is to build and validate the algorithm. Yet, a detailed exploration of the candidate motifs and the impact of the hyperparameters (fuzziness, density, size, etc.) is left for future work. 

The algorithms introduced here are general and the field of subgraphs mining is still rapidly evolving. We believe \vern\   could also be applied to other sources of data such as chemical compounds, protein networks, and gene expression networks to automatically mine for novel generalized structural patterns.

\section*{Implementation}
The source code is available at \url{vernal.cs.mcgill.ca}. We also provide a flexible interface and a user-friendly webserver to browse and download our results.

\section*{Acknowledgements}

The authors thank Vladimir Reinharz, Yann Ponty, Roman S. Gendron and Jacques Boitreaud for advice and support.

C.O. was funded by a PhD scholarship from Fonds de Recherche du Qu\'ebec Nature et technologies.
V. M. is funded by the INCEPTION project [PIA/ANR-16-CONV-0005] and benefits from support from the CRI through "Ecole Doctorale FIRE – Programme Bettencourt".
J.W. is supported by a Discovery grant from the Natural Sciences and Engineering Research Council of Canada.

\bibliographystyle{splncs04}
\bibliography{biblio}

\clearpage


\appendix
\setcounter{figure}{0} \renewcommand{\thefigure}{A.\arabic{figure}}

\setcounter{table}{0} \renewcommand{\thetable}{A.\arabic{table}}

\section*{Supplementary Material}

\section{RNA data}
\subsection{Chopping algorithm}
\label{supp:sect:chop}

We present here the algorithm used to chop RNA into fixed maximal size pieces. The idea of the algorithm is to recursively cut the RNA in halves up until the maximum size is reached. To minimally disrupt the structure, we cut the structure in the orthogonal direction to the principal axis of variation, according to Principal Component Analysis. This is detailed in \textbf{Algorithm \ref{supp:algo:chopper}}.

\begin{algorithm}
\SetKwFunction{printlcs}{\textsc{Print}-LCS}
\SetAlgoLined
\KwData{Full RNA Graph $g$, Maximum number of nodes $N$}
\KwResult{List of sub-structures of maximum size $N$.}

\If {$|g| \leq N$}
\Return $g$ \\
\Else{
    $g_a, g_b \leftarrow \text{ Split g in halves based on the PCA}$\\ 
      axes \\
    \Return chopper($g_a$)\\
    \Return chopper($g_b$)
}
 \caption{Chopper algorithm}
 \label{supp:algo:chopper}
\end{algorithm}

\subsection{Isostericity}
\label{supp:sect:isostericity}
\begin{figure}
	\centering
	\includegraphics[width=0.7\textwidth]{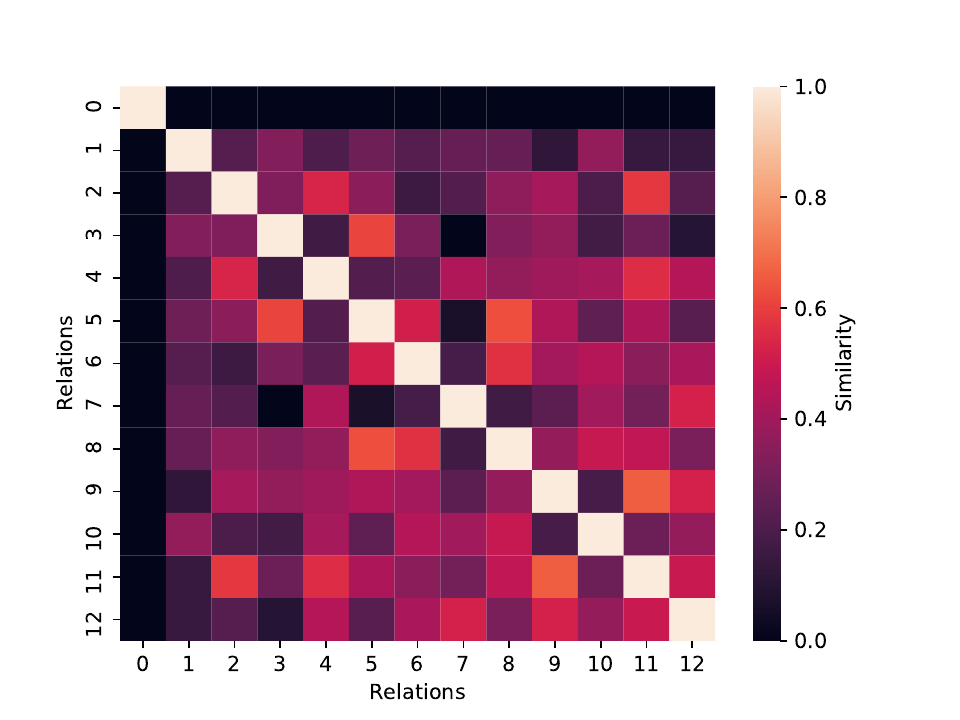}
	\caption{Isostericity matrix between relation types. }
	\label{supp:fig:isostericity}
\end{figure}

\fix{We represent RNA with graphs whose nodes are nucleotides and edges represent an interaction between those nucleotides. 
Let $\mathcal{R}$ denote the set of these interactions, this set contains backbone covalent interactions, 'cWW' interactions that are denoted as canonical base pairs and 11 other interactions that represent other geometries.}
Isostericity denotes the closeness between relation types at the 3-D level (base pairing geometries).
Stombaugh \cite{stombaugh2009frequency} computed the geometric discrepancy between all pairs of relation types (Shown in {\bf Figure \ref{supp:fig:isostericity}}).
We include a representation of the notion of isostericity between edge types in \textbf{Figure \ref{supp:fig:isostericity}}.

\fix{Graphs in RNA are directed but reciprocal ie if nodes A and B are linked by an edge A=(cSH)=>B, then automatically B and A are linked with an edge : B=(cHS)=>A.
The aforementioned isostericity relationship is symmetric, which means that for instance $\forall r \in \mathcal{R}, iso(cHS, r)= iso(cSH,r)$.
This would make our directed graphs as expressive as undirected graphs. 
A first solution is to simply make our graphs undirected and use the isostericity directly. 
However, relevant information is encoded in the direction of the edges, for instance the direction of the backbone.
Therefore, we expand the isostericity values by adding the following rules :
$$
\begin{cases}
    iso(b53,b35)=0.2 \\
    \forall r \in \mathcal{R}\setminus\{b53,b35\}, \ iso(b53,r)= 0  \\
    \forall r \in \mathcal{R}, \ iso(r,r)= 1  .
\end{cases}
$$
For all the other cases, we use the value reported in the isostericity matrix.
}

\section{Graph Edit Distance}
\label{supp:sec:ged}

\subsection{Model Selection with Graph Edit Distance}

The choice of similarity function $s_G$ is application specific.
One can use a function which maximizes performance on a downstream supervised learning task, or one can choose a similarity function which best encodes structural identity \cite{hamilton2017representation}.
Since supervised learning data for RNA 3D structures is scarce, we opt for the latter and  propose the Graph Edit Distance (GED) (or its similarity analog $exp[-GED]$) between rooted subgraphs, as this is the widely accepted yet computationally intensive gold standard for structure comparison  \cite{gao2010survey}.
Hence, we choose the $s_G$ which most closely correlates with GED.
Interestingly, GED is a generalization of the subgraph isomorphism problem \cite{bunke2008graph} which is at the core of previous RNA motif works such as \carna\  and \rthreem.

\subsection{Rooted GED}

In a nutshell, the GED between two graphs $G$, $H$ is the minimum cost set of modifications that can be made to $G$ in order to make it isomorphic to $H$.
This naturally encodes a notion of similarity since similar pairs will require few and inexpensive modifications, and vice versa.
The Graph Edit Distance (GED) between two graphs $g$ and $h$ is thus defined as follows:

\begin{equation}
GED(G, H) = \min_{(e_1,...,e_k) \in \Upsilon (G, H)} \sum_{i=1}^{k} c(o_i).
\end{equation}

where $\Upsilon$ is the set of all edit sequences which transform $G$ into $H$. 
Edit operations include: node/edge matching, deletion, and insertion. 
$c(o)$ is the cost of performing edit operation $o$ and $c$ is known as the cost function. 
Since we will be decomposing our graphs as rooted subgraphs, we define a slight modification to the GED formulation which compares two graphs given that their respective roots must be matched to each other. 
This algorithm is detailed in \textbf{Algorithm \ref{supp:algo:ged}}.

\begin{algorithm}
\SetAlgoLined
\KwData{
\begin{itemize}
    \item Pair of graphs $G$, $H$, (WLOG let $G$ be the smaller of the two graphs.)
    \item cost function $c$ 
    \item  heuristic $h$. 
    \item $r_G \in \mathcal{N}(G)$ root in first graph
    \item $r_H \in \mathcal{N}(G)$ root in second graph
\end{itemize}
  }
\KwResult{Minimum cost rooted distance and alignment between two graphs.}
$OPEN \leftarrow priorityQueue()$ \\
$V_{G} \leftarrow G.nodes()$\\
$V_{H} \leftarrow H.nodes()$\\
$v \leftarrow \text{first node in G}$
$OPEN.add((r_G,r_H'), c(r_G, r_H) + h(v_0, v')$
 \While{OPEN}{
	$v_{min} \leftarrow OPEN.pop()$\\
	Let $\mathcal{M}_{k} \leftarrow$ be partial mapping $\{(v_1, v'_1), .., (v_k, v'_k)\}$\\
	\If{$\vert \mathcal{M}_{k} \vert = \vert V_{G} \vert$}{
		\text{Mapping complete}\\
		\Return $\mathcal{M}_k$
	}
	\text{Add nodes at next depth}\\
	\ForEach{$u \in V_{H} \setminus v_{min}$}{
		$OPEN.add(v_{min} \cup (v_{k+1}, u), c(v_{k+1},u) + h(v_{k+1},u))$
	}
 }
 \caption{Rooted A* GED}
 \label{supp:algo:ged}
\end{algorithm}

\subsection{RNA GED}

We have adapted this algorithm to RNA data. 
We use the isostericity matrix ~\cite{stombaugh2009frequency} for edge substitutions, and do not apply a penalty to node substitutions.
Let $\mathcal{E}(.)$ be a function that returns the edge label for a given edge, and ISO the isostericity function which returns the similarity between edge types.
We define an RNA cost function over pair of edges $p$ and $q$ as follows :

\[ c(p \rightarrow q) = \text{ISO}(\mathcal{E}(p) ,\mathcal{E}(q))
\]
\[ c(p \rightarrow \emptyset) = \begin{cases}
    \alpha  & \text{backbone}\\
    \beta  & \text{canonical} \\
    \theta  & \text{non-canonical}
    \end{cases}
\]
For our experiments, we set $\alpha=1$, $\beta=2$, $\theta=3$ to emphasize the differences in non-canonical interactions between graphs.
We propose a simple modification to allow for comparison of rooted graphs ({\bf Algorithm ~\ref{supp:algo:ged}}), and use the general version of GED  to validate the ultimate full subgraph-level quality of our identified motifs.

\begin{figure}
\centering
\begin{minipage}{.35\textwidth}
\centering
    \includegraphics[width=\textwidth]{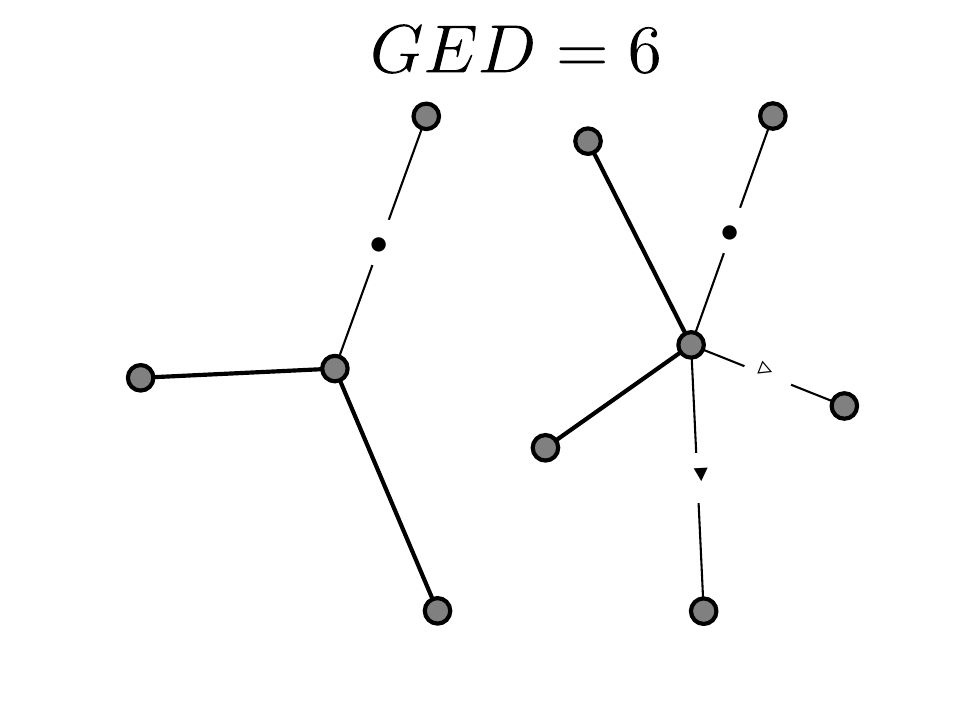}
    \label{fig:ged1}
\end{minipage}%
\begin{minipage}{0.35\textwidth}
\centering
    \includegraphics[width=\textwidth]{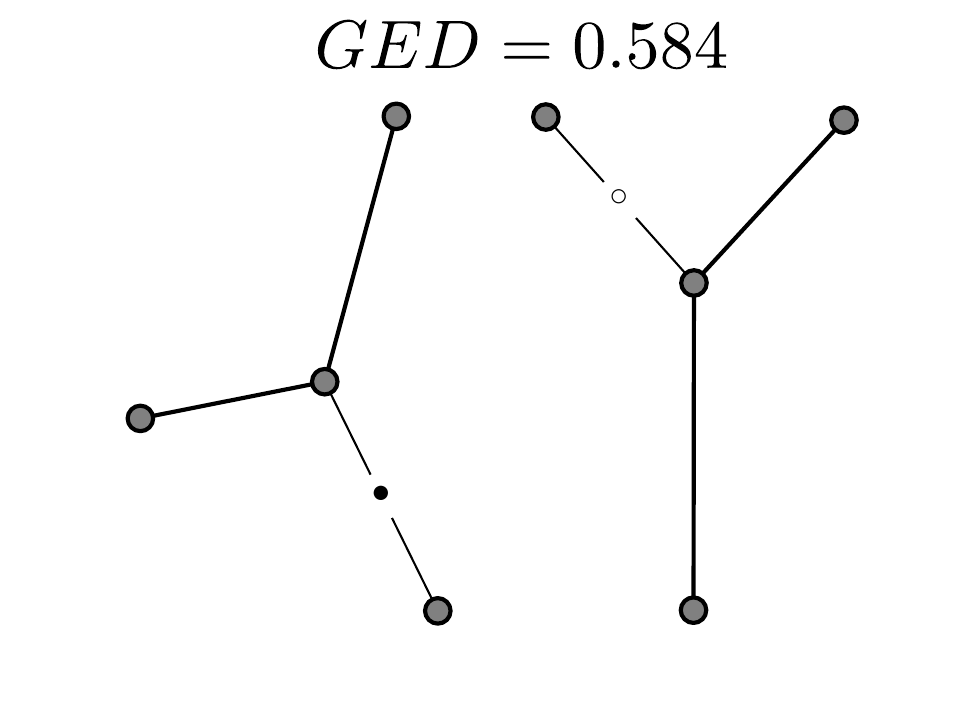}
    \label{fig:ged2}
\end{minipage}
\caption{Examples of similar and dissimilar pairs according to the GED A* algorithm.}
\label{supp:fig:ged}
\end{figure}

We include in \textbf{Figure \ref{supp:fig:ged}} an example of GED values for two pairs of graphlets, illustrating how similar graphs get lower values of distance.

\section{Rooted Subgraph Comparisons}
\label{supp:sect:kernels}

Here, we define a similarity function between a pair of rooted subgraphs, $g_u$ and $g_{u'}$.
We propose two main classes of $s_G$: ring-based, and matching-based similarity functions.

\subsection{Edge ring similarity functions}

The first $s_G$ functions we consider are a weighted sum of a distance between $l$-hop neighborhoods (aka rings) of each subgraph.
This formulation inspired by the function proposed in ~\cite{ribeiro2017struc2vec}.
We let $R^l_u = \{(u',w) : \delta(u,u') = l \quad \forall (u',w) \in E\}$ be the set of edges at distance $l$ from the root node $u$.
Let d be a normalized similarity function between two sets of edges. 
One such function can be formulated as a matching operation between the edges of each ring.
Let $\mb{X}$ is a binary matrix describing a matching from one ring to another and $\mb{S}$ a scoring function corresponding to this matching, d writes as :

\begin{align*}
    d(R_u, R_v) := \min -\sum_{e \in R_u} \sum_{e' \in R_{v}}\mb{S}_{\mathcal{L}(e), \mathcal{L}(e')} \mb{X}_{e,e'}
\end{align*}

Let $0 < \lambda < 1$ be a decay factor to assign higher weight to rings closer to the root nodes, and $N^{-1}$ be a normalization constant to ensure the function saturates at 1.
Then we can obtain a structural similarity for the rooted subgraphs around $u$ and $v$ as :

\begin{align*}
    k_L(u,v) := 1 - N^{-1} \sum_{l=0}^{L-1} \lambda^{l} d(R_u^l, R_v^l) \\
\end{align*}

The first function (\texttt{R\_1}) simply uses a delta function to compare to different edges. 
This assignment problem thus reduces to computing the intersection over union score between the histograms $f_R$ of edge labels found at each ring. 
However, this function treats all edge types equally and ignore the isostericity relationships.
The second function (\texttt{R\_iso}) has a matching value of 1 for backbone edges matched with backbone edges, 0 for backbone matched with any non covalent bond and the isostericity value for the similarity value of two non covalent bond. 

\subsection{Matching-based similarity functions}

Here, $s_G$ operates on the output of a function $f: g_u \rightarrow \Omega$ which decomposes a rooted subgraph into a set of objects $\Omega$ (e.g. sets/rings of nodes, edges, or smaller subgraphs such as graphlets \cite{shervashidze2009efficient}).
These objects can then be assigned structural and positional compatibilities.
We let  $\mb{C}_{\omega, \omega'}$ be the structural compatibility between objects $\omega, \omega'$, for example, edge isostericity.
Next, $\mb{D}_{\omega, \omega'}$ assigns a cost on pairs of objects depending on the relative path distance to their respective root nodes.
We propose various similarity functions, based on optimal matching of these objects with the most general form being:
\begin{equation}
    s_G(g_u, g_{u'}) := \min_{\mb{X}} \sum_{\omega \in \Omega} \sum_{\omega' \in \Omega'} (\alpha \mb{C}_{\omega, \omega'} + \beta \mb{D}_{\omega,\omega'}) \mb{X}_{\omega,\omega'}
    \label{eq:hung}
\end{equation}
where $\mb{X}$ is a binary matrix describing a matching from the elements of $\Omega$ to $\Omega'$, $\alpha$ and $\beta$ are user-defined weights for emphasizing positional vs structural compatibility.
We solve for the optimal matching between two sets of structural objects using the Hungarian algorithm \cite{Kuhn55thehungarian}.

Since the degree of our graphs is strongly bounded (max degree 5), we can define a graphlet as a rooted subgraph of radius 1 and obtain a manageable number of possible graphlets. 
This lets us define an $f$ which produces structurally rich objects.
Moreover the rooted aspect and the small size of those graphs make the GED computation tractable.

While the GED computation is tractable for such small graphs, it is still expensive when repeated many times.
For this reason, we implement a solution caching strategy which stores the computed GED when it sees a new pair of graphlets, and looks up stored solutions when it recognizes a previously seen pair \textbf{Supplementary Algorithm \ref{supp:algo:hash}}.

We can now define $\mb{S}$ from Equation \ref{eq:hung} as $\mb{S}_{ij} = \textrm{exp}[{-\gamma \textrm{GED}(g_i, g_j)]}$. We apply an exponential to the distance to bring the distances to the range [0,1], and convert them to a similarity. An optional scaling parameter $\gamma$ is included to control the similarity penalty on more dissimilar graphs. We also note that the construction of $\mb{S}$ can be parallelized but we leave the implementation for future work. 

\subsection{Additional settings}
We have experimented with several additional parameters. We tried including an Inverse Document Frequency (IDF) weighting to account for the higher frequency of non canonical interaction. This amounted to scale all comparison value by the product of the IDF term they involved.  

We also tried adding a re-normalization scheme to give higher values to matches of long rings. In particular, we want to express that a having a match of 9 out of 10 elements is stronger than having a match of 2 out of 3. Let $S$ be the raw matching score, $\mathbb{S}$ the normalized one and $L$ be the length of the sequences, we have tried two normalization settings, the \say{sqrt} and \say{log} ones : 
\begin{align*}
    \text{sqrt : } \mathbb{S} &= \big[ \frac{S}{L} \big] ^{\frac{5}{\sqrt{L}}}\\
    \text{log : }    \mathbb{S} &=  \big[ \frac{S}{L} \big] ^{\frac{1}{1+\log{L}}}\\
\end{align*}

\subsection{Graphlets hashing and distributions}
\begin{figure}
    \centering
    \begin{subfigure}[t]{0.4\textwidth}
        \centering
        \includegraphics[width=\textwidth]{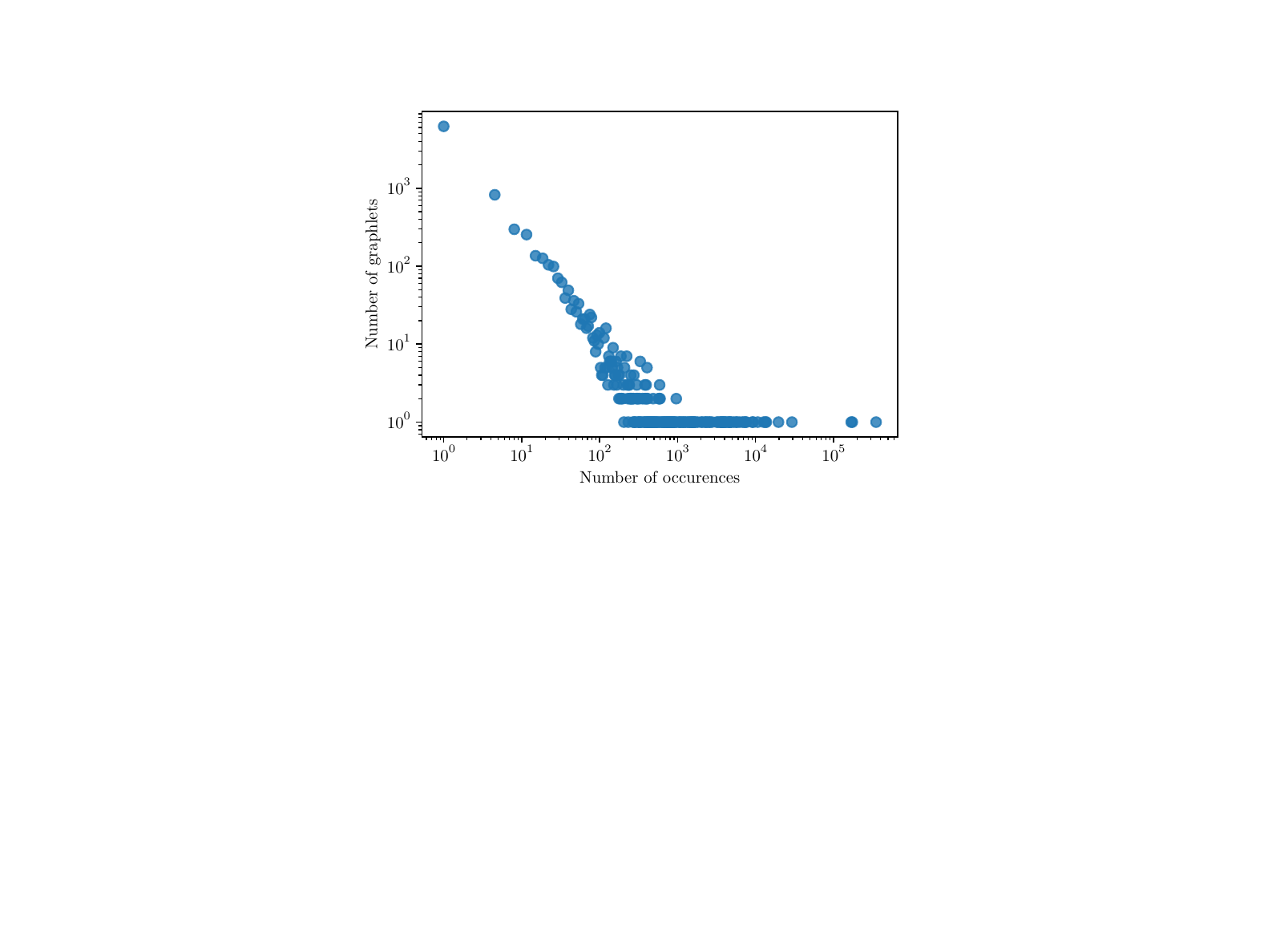}
        \caption{Graphlet frequency distribution}
    \end{subfigure}
    \begin{subfigure}[t]{.25\textwidth}
        \centering
        \includegraphics[width=\textwidth]{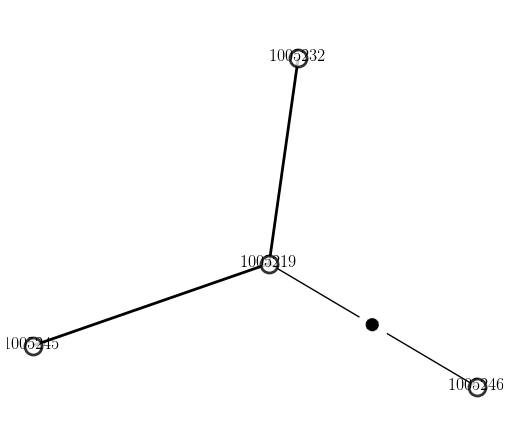}
        \caption{Most frequent graphlet}
    \end{subfigure}%
    \hfill
    \begin{subfigure}[t]{0.25\textwidth}
        \centering
        \includegraphics[width=\textwidth]{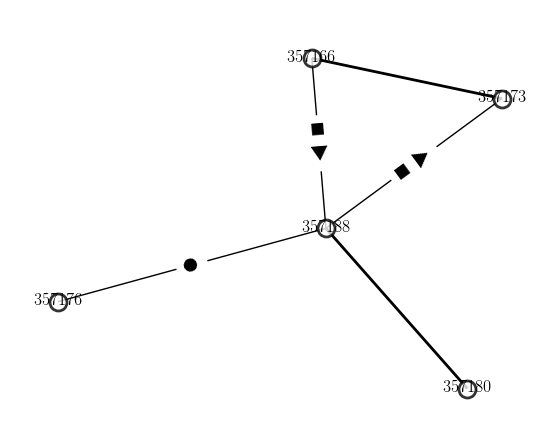}
        \caption{Example of a rare graphlet.}
    \end{subfigure}
    \hfill
    \caption{Graphlet distribution and examples.}
    \label{supp:fig:graphlets}
\end{figure}

We build a hash function which maps isomorphic graphlets to the same output, while assigning different outputs to non-isomorphic ones, allowing us to look up graphlet GED values.
This is done by building a sparse representation of an explicit Weisfeiler-Lehman isomorphism kernel, with a twist that edge labels are included in the neighborhood aggregation step.
The resulting hash consists of counts over the whole graphlet of hashed observed sequences of edge labels. We enforce the edge label hashing function to be permutation invariant by sorting the observed label sequence. In this manner, isomorphic graphs are given identical hash values regardless of node ordering.
Our hashing procedure outlined in {\bf Supplemental Algorithm \ref{supp:algo:hash}} also allows us to study the distribution of graphlets composing RNA networks {\bf Supplemental Fig \ref{supp:fig:graphlets}}, where we can observe a characteristic power law distribution.

\begin{algorithm}
\SetAlgoLined
\KwData{\begin{itemize}
	\item Graphlet $g$, 
	\item Maximum depth $K$
	\item $\textrm{HASH}$, function from strings to integers
	\item $\mathcal{L}$ function returning the label for an edge
	\end{itemize}}
\KwResult{Hash code for graphlet $h$}
$h \leftarrow counter()$\\
\ForEach{$k \in  \{1, .., K\}$}{
	\ForEach {$u \in g_N $}{
			$l_u^k \leftarrow \textrm{HASH}( \{\mathcal{L}(u,v) \oplus l_v^{k-1} \quad \forall v \in \mathcal{N}(u)\})$
	}
	$h \leftarrow h \cup counter(\{ l_u \quad \forall u \in g\})$
    }
\Return $h$
\caption{Weisfeiler-Lehman Edge Graphlet Hashing}
\label{supp:algo:hash}
\end{algorithm}

\section{Model Training}
\label{supp:sect:rgcn}

\subsection{Mathematical framing}

In order to identify related groups of rooted subgraphs using only the similarity function we would have to perform and store $N^2$ operations (where $N$ is the total number of nodes in $\mathbb{G}$.
When working with an order of $10^7$ nodes, this quickly becomes prohibitive. 
Once nodes in each graph are embedded into a vector space, searches and comparisons are much cheaper.

We therefore approximate the $s_G$ function over all pairs of rooted subgraphs using node embeddings.
We use a Relational Graph Convolutional Network (RGCN) model  \cite{schlichtkrull2018modeling} as parametric node embedding function $\phi(u) \rightarrow \mathbb{R}^d$ which maps nodes to a vector space of dimensionality $d$. 
The network is implemented in Pytorch \cite{paszke2017automatic} and DGL \cite{wang2019deep}. 
It is trained to minimize :

\begin{equation}
    \mathcal{L} = \| \langle \phi(g_u), \phi(g_{u'}) \rangle - s_G(g_u, g_{u'}) \|^2_2,
     \label{eq:loss}
\end{equation}

To focus performance on subgraphs that contain non canonical nodes and avoid the loss to be flooded by the canonical interactions (Watson Crick pairs), we then scale this loss based on the presence of non canonical interactions in the neighborhood of each node being compared. 
Given the rate of non canonical interactions $r$ and $\mathds{1}_{u}$ an indicator function that denotes the presence of non-canonical interactions in the neighborhood of node $u$, the scaling $\mb{S}_{u,v}$ of the $g_u, g_{u'}$ term writes as :

\begin{equation}
    \mb{S}_{g_u,g_{u'}} = (1+\frac{\mathds{1}_{g_u}}{r})(1+\frac{\mathds{1}_{g_{u'}}}{r}), \quad \mathcal{L}_{scaled} = \mb{S} \odot \mathcal{L}
\end{equation}

\subsection{Architecture and hyperparameters choices}
We use as many layers as the number of hops in the similarity function so that both functions have access to the same support in the subgraphs.
The embeddings resulting from this message passing are then fed to a Linear Layer.
The dimension per layers that we used were $[16, 32, 32]$, we used the default instanciation of DGL with ReLus and self-loops. 
We tried using Dropout but chose to not include it in our model in the end.
The model was then trained using Adam on sampling pairs on the order of k*N, where N is the number of chunks and $k \sim 5$. Empirically, we had convergence for values around $k=3$.

We can then perform clustering in the embedding space using any linear clustering algorithm.
We denote the resulting clusters as 1-motifs as they represent the aforementioned structural blocks of RNA.

\section{Metrics on the structural clusters}
\label{supp:sect:clusts}

\begin{figure}
    \centering
    \begin{subfigure}[t]{0.24 \textwidth}
        \centering
        \includegraphics[width=\textwidth]{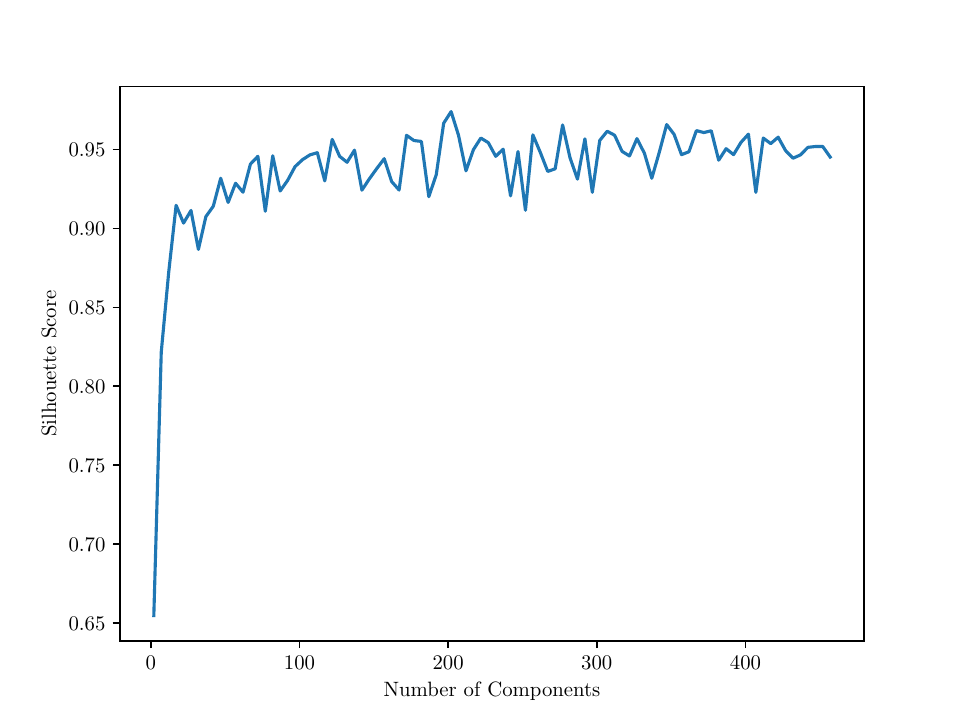}
        \caption{Silhouette score for different number of initial centroids.}
    \end{subfigure}%
    \begin{subfigure}[t]{0.24 \textwidth}
        \centering
        \includegraphics[width=\textwidth]{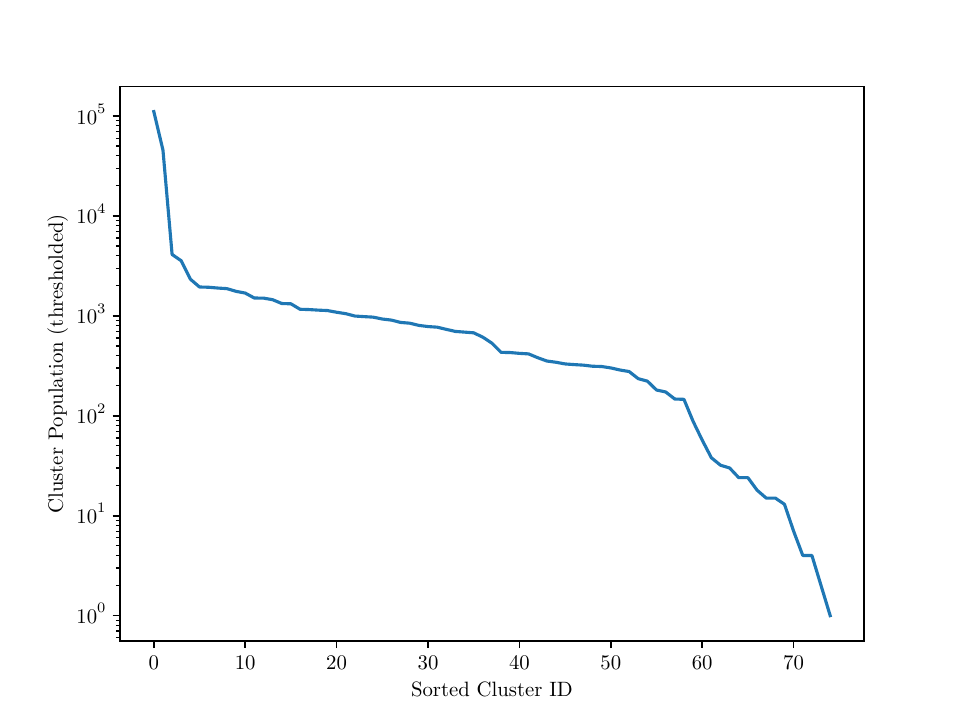}
        \caption{Nodes per cluster ids.}
    \end{subfigure}
    \begin{subfigure}[t]{0.24 \textwidth}
        \centering
        \includegraphics[width=\textwidth]{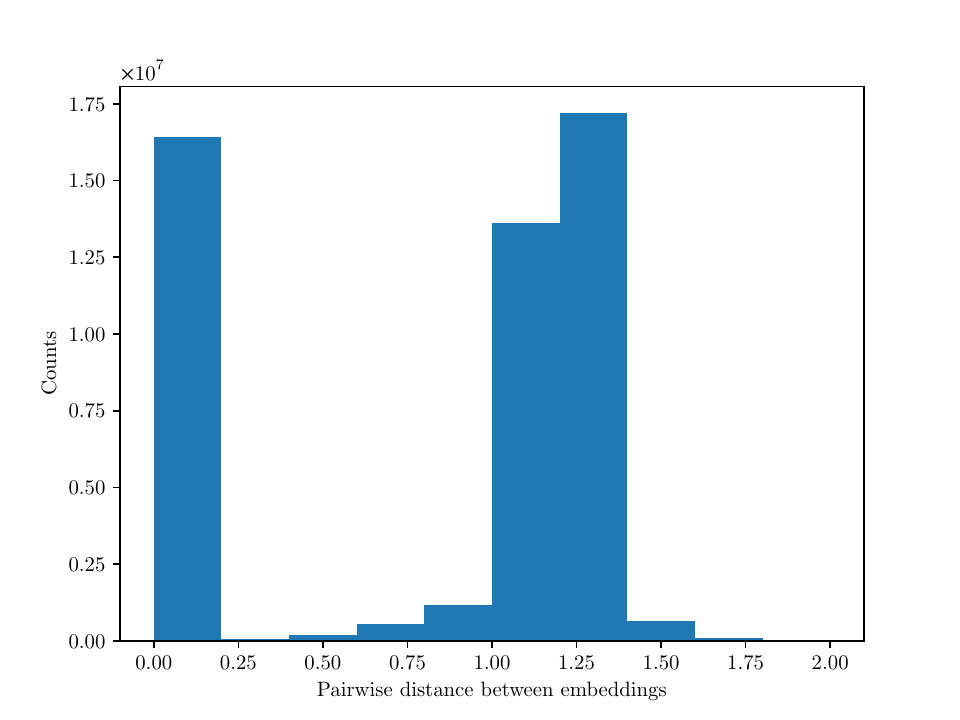}
        \caption{Distance between randomly selected pairs of embeddings.}
    \end{subfigure}
    \begin{subfigure}[t]{0.24 \textwidth}
        \centering
        \includegraphics[width=\textwidth]{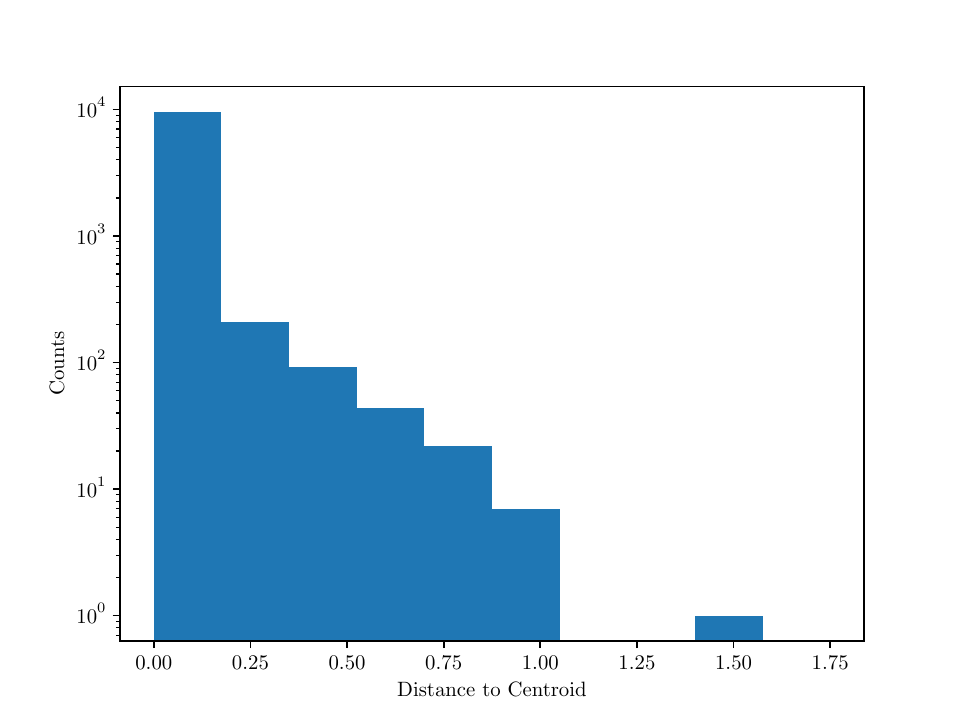}
        \caption{Distance between nodes embeddings and their centroid}
    \end{subfigure}
    \caption{Metrics when clustering with minibatch K-means, with $k=262$}
    \label{supp:fig:clustering}
\end{figure}

We present in \textbf{Figure \ref{supp:fig:clustering}} some metrics on the clustering of our structural embeddings. The silhouette scores suggest that the number of clusters necessary to obtain satisfying scores is around 60.
However seeding the method with more centroids can yield better results, even if they end up collapsed in the end. 
We used 200 centroids that got collapsed into 60 distinct centroids.

The cluster population have an interesting distribution, with the two first clusters heavily populated (stem nodes), then an interesting population of graphlets and finally very rare ones.
Despite the curse of dimensionality, we see that the nodes are much closer to their centroids than they are one from another.
These metrics suggest that the structures present in RNA do fall into well separated clusters.

\section{Retrieve complexity}
\label{supp:sect:complexity}
In this section, we want to investigate the complexity of the retrieve algorithm. This complexity depends highly on both the topology of the meta-graph, the query graph and the individual RNA-graphs.

Let $N_p$ be the number of edges in each of the $p$ parallel edges of the query graph, and $E_p$ be the corresponding meta-edge. 
We consider the edges of the query graph in the order of the parallel edges, let $p(t) = \min_k, \sum_k N_k > t$, the parallel edge considered at time t.

At each step t, the complexity bound is going to depend on the number of candidate motifs inside each RNA graphs at time t-1 as well as the number of possible additional edges to insert into those candidates. If we denote as $M_{g,t}$ the number of candidates in graph g at time t, and $N_{g,t}$ the number of edges in graph g that belong to the meta-edge $E_{p(t),g}$, the complexity writes as $O(\sum_t \sum_{g \in \mathcal{G}} E_{p(t),g} M_{g,t})$
The term $E_{p(t),g}$ mostly acts as a sparsity term, as it would not exceed ten but can very often be zero if the graph does not include such an edge. Therefore, we introduce the notation $\mathcal{G}_p$, the set of RNA graphs that contain an edge in $E_p$, to omit this term.
We also introduce $t_{g,t} = \sum_{l<t} E_{p(l),g}$ the number of query edges explored present in graph g at time t.

Let us now try to address the second term $M_{g,t}$.
We divide our algorithm into each of the parallel edges and dive into the evolution of this term.
$M_{g,t}$ represents the number of combinations of nodes in a given graph that are currently considered as a candidate motif.
Every different p is going to launch a combinatorial explosion that results from the numerous possibilities of combining nodes.
For RNA graphs, it is mostly a problem for stems that are ubiquitous and all share the same structure. Starting from all stem nodes, after one merging step we have to add all possible combinations of adjacent stem nodes. 
We give a loose bound of this number that is practical for our application, but note that after sufficient merging, there are just the full stems as candidates, showing that this bound becomes loose.

We rely on the fact that we can delete a partial solution if it is completely contained into another one because its expansion can only result in lower scores. Thus after a given motif got expanded we can remove it from $\mathcal{M}$. 
Therefore, we can bound this number by counting the number of children a given element can produce. Using sets structure enables fast neighborhood checking operations but also ensures we do not add the same object twice.
For stems that are the worst case scenario, each connected component of length $k$ can yield a maximum of $k+2$ children but any of this children is added twice because the edge has two ends. 
Therefore after k expansions,we have $M_{g,t+k} < M_{g,t} \frac{(k+2)!}{2^k}$. 
This results in a final loose complexity bound of : 
$O(\sum_t \sum_{g \in \mathcal{G}_{p(t)}} \frac{(t_{g,t}+2)!}{2^{t_{g,t}}} )$.

As mentioned before this is not a good bound when $k$ grows because the stems get completed. Therefore, there are a lot of ways to select 3 adjacent stem nodes of a 6 stem, but only one to get all 6. 
In practice, during each cycle, the number $M_{g,t}$ grows and decreases, making the algorithm tractable, starting back from reasonable numbers at each cycles. Empirically, running the retrieve algorithm on a single core rarely exceed a few seconds.

\section{MAA}
\label{supp:sect:maa}
\vspace{0pt}

\begingroup
\makeatletter
\let\@latex@error\@gobble
\makeatother
\begin{algorithm}
\SetAlgoLined
\KwData{\begin{itemize}
	\item Meta-Graph $\mathcal{G}$, 
    \item Minimum density $\delta$ (number of motif instances)
    \item Number of steps $K$
	\end{itemize}}
\KwResult{List of meta-graphs} 
$\mathcal{M} \leftarrow list() $ \\
$\mathcal{E} \leftarrow \mathcal{G}.edges() $\\
\ForEach{$k \in  \{1, .., K\}$}{
   $\mathcal{E}' \leftarrow \varnothing$ \\
   $\mathcal{M}[k] \leftarrow list()$ \\
   \While{$ \mathcal{E}$}{
    $m, m' \leftarrow \mathcal{E}.pop()$ \\
    $\mu \leftarrow \texttt{merge}(m.subgraphs, (m, m'))$\\
    \If{$\vert \mu \vert > \delta $}{
        $\mathcal{M}[t].append(\mu)$\\
        \tcc{Connect new node to adjacent clusters}
         \ForEach{$c' \in \mathcal{N}(\mu)$}{
                 $\mathcal{E}'.add((\mu,c'))$
            }
     }
    $\mathcal{E} \leftarrow \mathcal{E}'$
    }}
    \Return $\mathcal{M}$
    \caption{{\bf M}otif {\bf A}ggregation {\bf A}lgorithm (MAA).  At each step, $k$, the algorithm iterates through edges $(m, m')$ of the meta-graph, applying Algorithm ~\ref{algo:merging} to construct a $k+1$ motif $\mu$.
 The updated meta-connectivity is stored as new meta-edges. }
\label{algo:motif} 
\end{algorithm}
\endgroup

\section{Full results for the similarity function validations}

We include in this section the full results we got for a grid search validation in the absence of a better way to guide our intuition {\bf Table ~\ref{supp:table:onehop}, ~\ref{supp:table:twohop}}, as well as a condensed version in {\bf Table ~\ref{table:main_ged_corr}}.

\begin{table*}
\centering
\begin{tabular}{lrrl|rrrrr}
    \toprule 
     {\bf method} &  {\bf depth} &  {\bf decay} & {\bf normalization} & 
     {\bf r} &  {\bf r\_th} &  {\bf time (s)} \\
     \midrule
edge hist. &      1 &  0.5 & sqrt      & 0.69         & 0.80         & \textbf{$<$0.001} \\
edge hist. + iso & 1 &  0.5 & sqrt     & 0.74         & 0.80         & \textbf{$<$0.001} \\
edge hungarian + iso & 1 &  $-$ & sqrt & 0.80         & 0.91         & \textbf{$<$0.001} \\
graphlets hist. &    1 &  any & None   & \textbf{1.0} & \textbf{1.0} & 0.029 \\
graphlet hungarian &  1 &   $-$ & None & \textbf{1.0} & \textbf{1.0} & 0.030 \\
edge hungarian + iso & 2 &  $-$ & sqrt & \textbf{0.64}& \textbf{0.75}& \textbf{$<$0.001} \\
graphlets hist. &     2 &  0.8 & sqrt  & 0.63         & 0.68         & 0.102 \\
graphlet hungarian &  2 &  $-$ & sqrt  & 0.63         & 0.70         & 0.433 \\
\midrule
1hop RGCN &      1 & $-$ & $-$ & 0.826 & 0.927 & \textbf{$<$0.001} \\
2hop RGCN &      2 & $-$ & $-$ & 0.596 & 0.577 & \textbf{$<$0.001} \\
    \bottomrule
\end{tabular}
\caption{Correlation with the GED for different kernels and embedding settings. For each experiment (row) we report the Pearson correlation coefficient $\mathbf{r}$. Correlations are computed pairwise on a subset of 200 nodes, as described in the main text. We include the result on this whole data as $\mathbf{r}$, as well as a subset which include only pairs of nodes with a GED below 6 ({\bf r\_th})
}
\label{table:main_ged_corr}
\end{table*}

\begin{table*}
\centering
\begin{tabular}{lrrlrr}
\toprule
     Method &  Decay &  IDF & Normalization &  Correlation &  Thresholded Correlation \\
\midrule
   graphlet &    NaN &  NaN &          None &     0.999982 &                 1.000000 \\
R\_graphlets &    0.8 &  NaN &          None &     0.999982 &                 1.000000 \\
R\_graphlets &    0.5 &  NaN &          None &     0.999982 &                 1.000000 \\
R\_graphlets &    0.3 &  NaN &          None &     0.999982 &                 1.000000 \\
   graphlet &    NaN &  NaN &          sqrt &     0.907263 &                 0.945552 \\
R\_graphlets &    0.8 &  NaN &          sqrt &     0.907263 &                 0.945552 \\
R\_graphlets &    0.5 &  NaN &          sqrt &     0.907263 &                 0.945552 \\
R\_graphlets &    0.3 &  NaN &          sqrt &     0.907263 &                 0.945552 \\
  hungarian &    NaN &  0.0 &          None &     0.796640 &                 0.910618 \\
  hungarian &    NaN &  0.0 &          sqrt &     0.784811 &                 0.916345 \\
  hungarian &    NaN &  1.0 &          sqrt &     0.777268 &                 0.916563 \\
  hungarian &    NaN &  1.0 &          None &     0.756889 &                 0.862990 \\
      R\_iso &    0.5 &  1.0 &          sqrt &     0.737533 &                 0.798435 \\
      R\_iso &    0.8 &  1.0 &          sqrt &     0.737533 &                 0.798435 \\
      R\_iso &    0.3 &  1.0 &          sqrt &     0.737533 &                 0.798435 \\
      R\_iso &    0.3 &  0.0 &          sqrt &     0.731762 &                 0.799086 \\
      R\_iso &    0.5 &  0.0 &          sqrt &     0.731762 &                 0.799086 \\
      R\_iso &    0.8 &  0.0 &          sqrt &     0.731762 &                 0.799086 \\
      R\_iso &    0.3 &  1.0 &          None &     0.715601 &                 0.817003 \\
      R\_iso &    0.8 &  1.0 &          None &     0.715601 &                 0.817003 \\
      R\_iso &    0.5 &  1.0 &          None &     0.715601 &                 0.817003 \\
      R\_iso &    0.5 &  0.0 &          None &     0.701981 &                 0.829820 \\
      R\_iso &    0.3 &  0.0 &          None &     0.701981 &                 0.829820 \\
      R\_iso &    0.8 &  0.0 &          None &     0.701981 &                 0.829820 \\
        R\_1 &    0.5 &  1.0 &          sqrt &     0.689727 &                 0.798827 \\
        R\_1 &    0.5 &  1.0 &          None &     0.689727 &                 0.798827 \\
        R\_1 &    0.3 &  1.0 &          sqrt &     0.689727 &                 0.798827 \\
        R\_1 &    0.3 &  1.0 &          None &     0.689727 &                 0.798827 \\
        R\_1 &    0.8 &  1.0 &          sqrt &     0.689727 &                 0.798827 \\
        R\_1 &    0.8 &  1.0 &          None &     0.689727 &                 0.798827 \\
        R\_1 &    0.3 &  0.0 &          sqrt &     0.683472 &                 0.838047 \\
        R\_1 &    0.3 &  0.0 &          None &     0.683472 &                 0.838047 \\
        R\_1 &    0.5 &  0.0 &          sqrt &     0.683472 &                 0.838047 \\
        R\_1 &    0.8 &  0.0 &          None &     0.683472 &                 0.838047 \\
        R\_1 &    0.5 &  0.0 &          None &     0.683472 &                 0.838047 \\
        R\_1 &    0.8 &  0.0 &          sqrt &     0.683472 &                 0.838047 \\
\bottomrule
\end{tabular}
\caption{One hop rooted subgraph correlation to GED}
\label{supp:table:onehop}
\end{table*}

\begin{table*}[p]
\centering
\begin{tabular}{lrrlrr}
\toprule
     Method &  Decay &  IDF & Normalization &  Correlation &  Thresholded Correlation \\
\midrule
  hungarian &    NaN &  1.0 &          sqrt &     0.640183 &                 0.745018 \\
   graphlet &    NaN &  NaN &          sqrt &     0.627242 &                 0.701199 \\
R\_graphlets &    0.8 &  NaN &          sqrt &     0.625675 &                 0.683274 \\
  hungarian &    NaN &  0.0 &          sqrt &     0.606648 &                 0.689525 \\
  hungarian &    NaN &  1.0 &          None &     0.589180 &                 0.624824 \\
R\_graphlets &    0.5 &  NaN &          sqrt &     0.588992 &                 0.598480 \\
R\_graphlets &    0.8 &  NaN &          None &     0.570733 &                 0.584230 \\
      R\_iso &    0.8 &  1.0 &          sqrt &     0.565505 &                 0.546167 \\
      R\_iso &    0.8 &  0.0 &          sqrt &     0.558772 &                 0.541444 \\
R\_graphlets &    0.3 &  NaN &          sqrt &     0.557002 &                 0.512982 \\
   graphlet &    NaN &  NaN &          None &     0.554304 &                 0.595573 \\
  hungarian &    NaN &  0.0 &          None &     0.552988 &                 0.664127 \\
R\_graphlets &    0.5 &  NaN &          None &     0.548846 &                 0.519385 \\
      R\_iso &    0.8 &  1.0 &          None &     0.538128 &                 0.532596 \\
      R\_iso &    0.5 &  1.0 &          sqrt &     0.529356 &                 0.499034 \\
R\_graphlets &    0.3 &  NaN &          None &     0.527732 &                 0.455359 \\
      R\_iso &    0.5 &  0.0 &          sqrt &     0.524007 &                 0.492044 \\
      R\_iso &    0.8 &  0.0 &          None &     0.512851 &                 0.516201 \\
      R\_iso &    0.5 &  1.0 &          None &     0.506934 &                 0.479951 \\
        R\_1 &    0.8 &  1.0 &          sqrt &     0.500248 &                 0.506578 \\
        R\_1 &    0.8 &  1.0 &          None &     0.500248 &                 0.506578 \\
      R\_iso &    0.3 &  1.0 &          sqrt &     0.491474 &                 0.439565 \\
      R\_iso &    0.3 &  0.0 &          sqrt &     0.487652 &                 0.432580 \\
      R\_iso &    0.5 &  0.0 &          None &     0.485761 &                 0.463078 \\
        R\_1 &    0.8 &  0.0 &          sqrt &     0.481469 &                 0.506201 \\
        R\_1 &    0.8 &  0.0 &          None &     0.481469 &                 0.506201 \\
      R\_iso &    0.3 &  1.0 &          None &     0.474367 &                 0.420955 \\
        R\_1 &    0.5 &  1.0 &          None &     0.469777 &                 0.453362 \\
        R\_1 &    0.5 &  1.0 &          sqrt &     0.469777 &                 0.453362 \\
      R\_iso &    0.3 &  0.0 &          None &     0.456932 &                 0.406642 \\
        R\_1 &    0.5 &  0.0 &          sqrt &     0.452601 &                 0.450131 \\
        R\_1 &    0.5 &  0.0 &          None &     0.452601 &                 0.450131 \\
        R\_1 &    0.3 &  1.0 &          sqrt &     0.438522 &                 0.395088 \\
        R\_1 &    0.3 &  1.0 &          None &     0.438522 &                 0.395088 \\
        R\_1 &    0.3 &  0.0 &          sqrt &     0.422765 &                 0.391077 \\
        R\_1 &    0.3 &  0.0 &          None &     0.422765 &                 0.391077 \\
\bottomrule
\end{tabular}
\caption{Two hop rooted subgraph correlation to GED.}
\label{supp:table:twohop}
\end{table*}

\section{MAA supplemental results}

\begin{table*} 
\centering
\begin{tabular}{lrr}
\toprule
 Motif Size &  \shortstack{Number \\ of Motifs} &  \shortstack{Mean Number\\ of Instances} \\
\midrule
          1 &        48 &      2041.90 \\
          2 &       101 &       830.22 \\
          3 &       163 &      1112.97 \\
          4 &       260 &      1090.64 \\
          5 &       460 &      1264.23 \\
          6 &       823 &      1318.79 \\
          7 &      1641 &      1308.76 \\
\bottomrule
\end{tabular}
\caption{Number of motifs and of instances per motif for each size, as found by the MAA algorithm}
\label{supp:table:sizes}
\end{table*}

\setlength{\tabcolsep}{18pt}
\begin{table}
\centering
\begin{tabular}{lrr}
\toprule
    Dataset &  Covered & Missed \\
\midrule
       BGSU \cite{petrov2013automated} &       112 &      14 \\
 RNA3DMotif \cite{djelloul2009algorithmes} &        2 &       0 \\
   CaRNAval \cite{reinharz2018mining} &       147 &      10 \\
\bottomrule
\end{tabular}
\vspace{0.5in}
\caption{Nearly all motifs identified by three published RNA motif tools are a subset of the motifs found by \vern.}
\label{table:hmp}
\end{table}

\begin{figure*} 
	\centering
	\includegraphics[width=\textwidth]{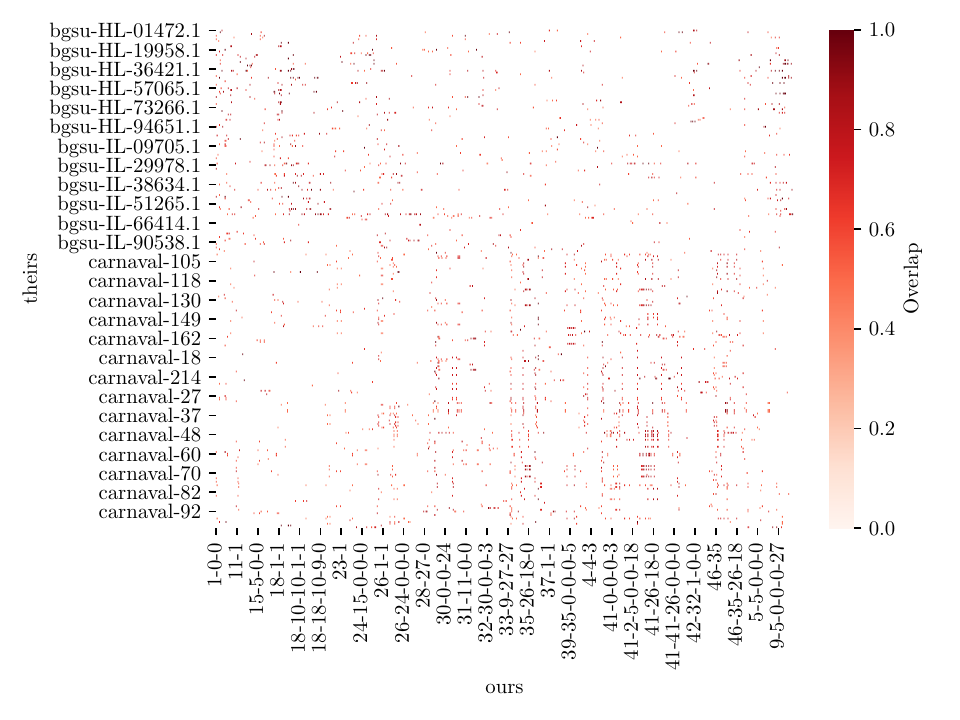}
	\caption{Agreement with existing motif libraries. Each cell value ranges from 0 to 1, where 1 indicates that all the nodes of an instance of a known motif are contained in a \vern\  motif.}
	\label{fig:overlap}
\end{figure*}

\begin{figure*} 
	\centering
	\includegraphics[width=\textwidth]{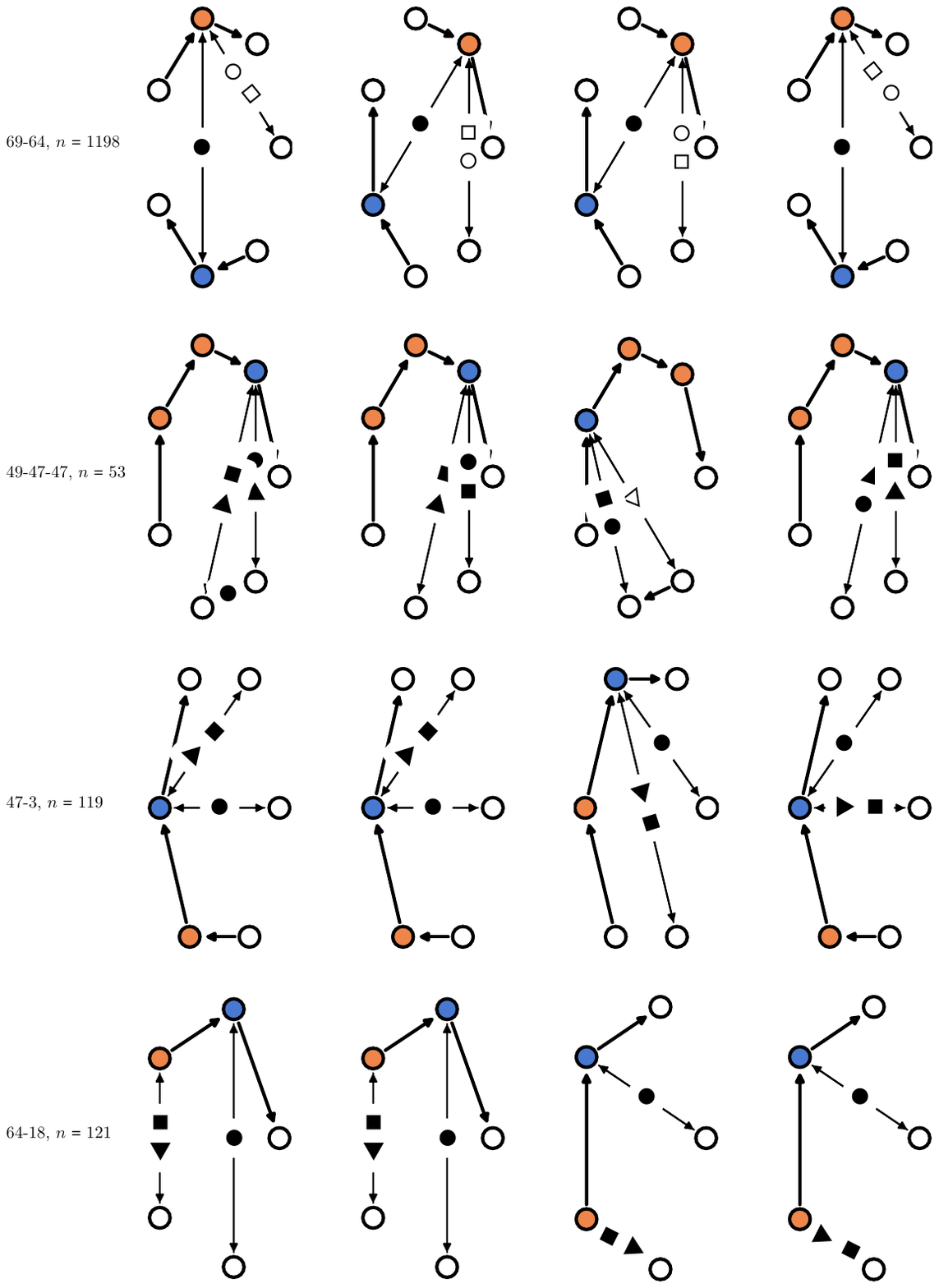}
	\caption{More example instances of motifs.}
	\label{fig:supp:sample-motifs}
\end{figure*}
\FloatBarrier
	
\end{document}